\documentclass[aps,10pt,twocolumn,amsmath,amssymb,showpacs,floatfix,longbibliography]{revtex4-2}

\usepackage{graphicx}
\usepackage[utf8]{inputenc}

\usepackage[label font={large}]{subfig}
\usepackage[export]{adjustbox}
\usepackage{rotating}
\usepackage{booktabs, makecell, tabularx}
\usepackage{color}
\usepackage{dcolumn} 
\usepackage{bm}
\usepackage{mathtools}
\usepackage{latexsym}
\usepackage{float}
\usepackage{amsfonts}
\usepackage{amsmath}   
\usepackage{amssymb}  
\usepackage{babel}
\usepackage{natbib}
\usepackage[colorlinks]{hyperref}
\hypersetup{ 
	colorlinks=true, 
	linkcolor=blue, 
	filecolor=blue, 
	citecolor = blue,       
	urlcolor=blue, 
}

\DeclareUnicodeCharacter{2212}{-}

\begin{document}

\title{Nonconventional screening of Coulomb interaction in two-dimensional semiconductors and metals: A comprehensive cRPA study of MX$_2$ (M=Mo, W, Nb, Ta; X=S, Se, Te)}

\author{H. R. Ramezani$^{1}$}
\author{E. \c{S}a\c{s}{\i}o\u{g}lu$^{2}$}\email{ersoy.sasioglu@physik.uni-halle.de}
\author{H. Hadipour$^{1}$}\email{hanifhadipour@gmail.com}
\author{H. Rahimpour Soleimani$^{1}$}
\author{C. Friedrich$^{3}$}
\author{S. Bl\"{u}gel$^{3}$}
\author{I. Mertig$^{2}$}	

\affiliation{$^{1}$Department of Physics, University of Guilan, 41335-1914, Rasht, Iran \\
$^{2}$Institute of Physics, Martin Luther University Halle-Wittenberg, 06120 Halle (Saale), Germany \\
$^{3}$Peter Gr\"unberg Institut, Forschungszentrum J\"ulich 
and JARA, 52425 J\"ulich, Germany}

\date{\today}

\begin{abstract}

Two-dimensional (2D) semiconducting and metallic transition metal dichalcogenides (TMDs) 
have attracted significant attention for their promising applications in a variety of fields. 
Experimental observations of large exciton binding energies and non-hydrogenic Rydberg series 
in 2D semiconducting TMDs, along with deviations in plasmon dispersion in 2D metallic TMDs, 
suggest the presence of a nonconventional screening of the Coulomb interaction. 
The experimentally observed Mott insulating state in the charge density wave (CDW) reconstructed 
lattice of TMDs containing $4d$ and $5d$ elements further confirms the presence of strong Coulomb 
interactions in these systems. In this study, we use first-principles electronic structure 
calculations and constrained random-phase approximation to calculate the Coulomb 
interaction parameters (partially screened $U$ and fully screened $W$) between localized 
$d$ electrons in 2D TMDs. We specifically explore materials represented by the formula 
MX$_2$ (M=Nb, Ta, Mo, W, and X=S, Se, Te) and consider three different phases ($1H$, $1T$, 
and $1T'$). Our results show that the short-range interactions are strongly screened in all three 
phases, whereas the long-range interactions remain significant even 
in metallic systems. This nonconventional screening provides a compelling explanation for the 
deviations observed in the usual hydrogenic Rydberg series and conventional plasmon dispersion 
in 2D semiconducting and metallic TMDs, respectively. Our calculations yield on-site Coulomb 
interaction parameters $U$ within the ranges of 0.8-2.5 eV, 0.8-1.9 eV, and 0.9-2.4 eV for the $1H$, $1T$, 
and $1T'$ structures, respectively. These values depend on the specific chalcogen X, the number 
of $d$ electrons, and the correlated subspace.  Using the calculated $U$ parameters for the
undistorted $1T$ structure, we extract the on-site effective $U^{\mathrm{eff}}_{00}$ and 
nearest neighbor $U^{\mathrm{eff}}_{01}$ Coulomb interaction parameters for reconstructed 
commensurate CDW NbX$_2$ and TaX$_2$ compounds. Furthermore, our findings indicate a substantially 
high ratio of on-site effective Coulomb interaction to bandwidth ($U^{\mathrm{eff}}_{00}/W_b \gg 1$) 
in CDW TMDs, providing robust evidence for the experimentally observed strongly correlated Mott phase.
This work sheds light on the nonconventional screening of Coulomb interactions in 2D TMDs, offering 
valuable insights into their electronic properties and potential applications in emerging technologies. 
It advances our fundamental understanding of these materials and holds promise for their use in various
applications.

\end{abstract}
		

\maketitle

\section{Introduction} \label{sec:Introduction}

Two-dimensional (2D) transition metal dichalcogenides (TMDs) have attracted significant 
attention for their unique electronic \cite{geim2021exploring,li2017functionalization,chen2022recent}, 
magnetic \cite{yang2023emerging,hossain2022synthesis}, optical \cite{kumbhakar2021advance,gong2017electronic,mueller2018exciton},  
and valleytronic properties \cite{norden2019giant,zhao2021valley,edwards2023giant,du2019giant}. 
Semiconducting TMDs have tunable band gaps \cite{Ramasubramaniam_2011,Sun_2023} and high charge 
carrier mobility, making them promising candidates for next-generation electronic 
devices that can overcome the limitations of traditional silicon-based technology 
\cite{Lembke_2015,Radisavljevic_2011}. TMDs also have strong light-matter interaction 
and valley-dependent electronic properties, making them ideal for developing valleytronic 
devices such as valley-selective photodetectors and light-emitting diodes \cite{liu2015strong,shafi2022inducing,Zeng_2018}.
In addition to their potential applications in optoelectronics and valleytronics, TMDs have also been shown to 
have promising thermoelectric properties \cite{balendhran2013two,zhang2017thermoelectric,kanahashi20202d,jia2022recent}. 
TMDs have a high thermoelectric figure of merit ($Z_T$), which is a measure of their 
efficiency as thermoelectric materials. This makes them promising candidates for 
applications in thermoelectric energy harvesting, waste heat recovery, and refrigeration.

\begin{figure}[!h]
\begin{center}
\includegraphics[scale=0.128]{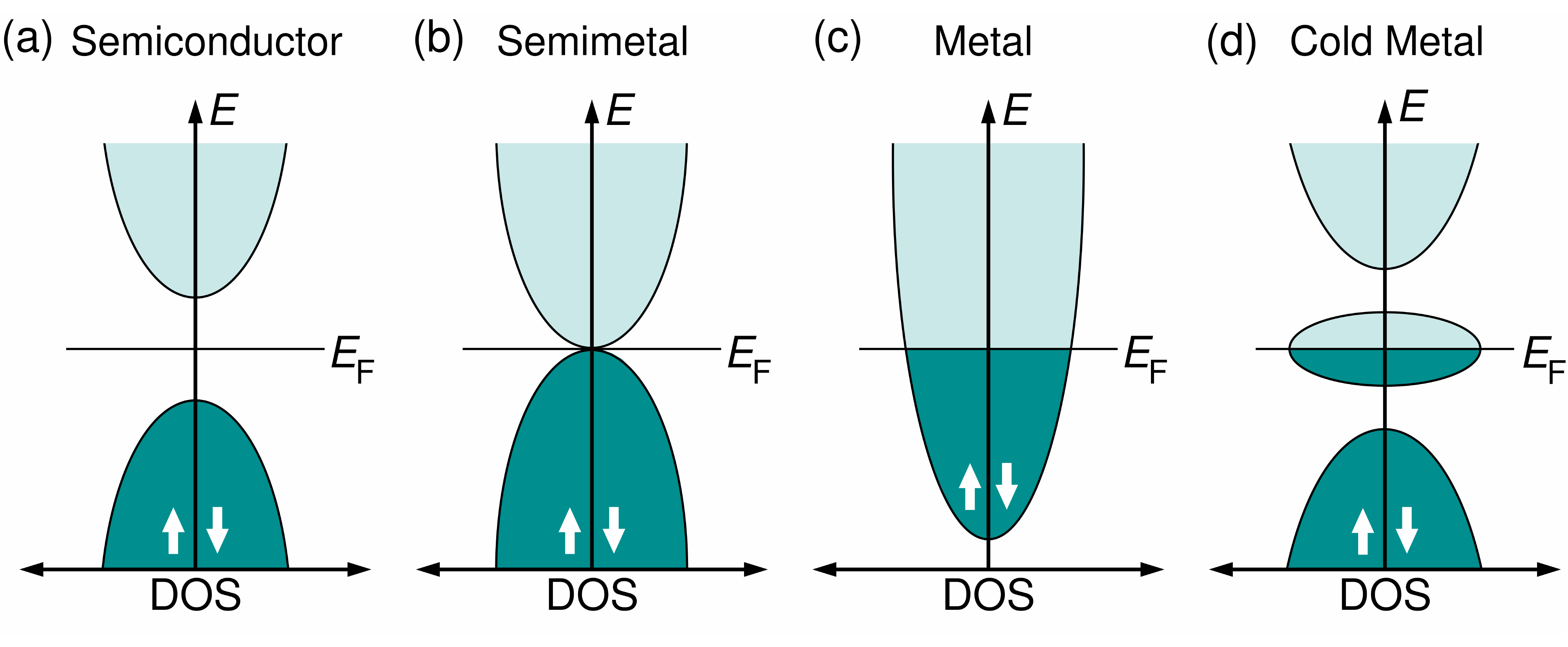}
\end{center}
\vspace*{-0.5cm} 
\caption{(Colors online) Schematic representation of the density of states  
for a semiconductor (a), a semimetal (b), a metal (c), and a cold metal (d).} 
\label{fig0}
\end{figure}

Metallic TMDs, such as "cold metals" like 1$H$- NbSe$_2$ (see Fig.\,\ref{fig0}), 
have opened up new possibilities for device innovation. One notable application 
is their potential to enable tunnel diodes exhibiting negative differential 
resistance with ultra-high peak-to-valley current ratios \cite{sasioglulu2023theoretical}.
Cold metallic TMDs have also been shown to achieve sub-60 mV/dec subthreshold swing 
in CMOS transistors, which could lead to significant improvements in transistor performance 
\cite{qiu2018dirac,liu2020switching,marin2020lateral,logoteta2020cold,tang2021steep}. In 
addition to their applications in nanoelectronics, metallic TMDs exhibit intriguing low-temperature 
properties such as unconventional plasmon dispersion \cite{Cudazzo_2016,da2020universal}, 
charge density wave (CDW) phenomena \cite{Ritschel_2015,chen2020strong,Lee_2021}, 
superconductivity \cite{Sipos_2008,Navarro_Moratalla_2016}, and magnetism 
\cite{Xu_2014,G_ller_2016,Pasquier_2022}. These diverse properties are governed by 
the crystal phases of MX$_2$, which can be trigonal prismatic $1H$, octahedral $1T$, or 
distorted octahedral $1T'$ (see Fig.\,\ref{fig:struct})  \cite{Chen_2022,Maghool_2023}. TMDs are therefore 
a promising platform for exploring a wide range of phenomena, with the potential to 
make significant contributions to both fundamental science and technological applications.

The screening of the Coulomb interaction in reduced dimensions is of fundamental interest 
for practical applications, as it affects the transport and optical properties of low-dimensional 
devices. In 2D materials containing TM atoms, the dielectric screening of Coulomb interactions is significantly 
reduced due to the confinement effect and presence of narrow $d$ electronic states \cite{Karbalaee_Aghaee_2022,yekta_2021}. 
This reduced screening has important consequences for the properties of semiconducting TMDs, 
such as the MoS$_2$ monolayer. For example, it leads to the formation of tightly bound excitons with 
large binding energies up to 1 eV \cite{ebe_1,ebe_4,WS_1,WS_2,MoSe,Kobayashi_2023,Chen_2023}.
Additionally, the exciton excitation spectra in TMDs monolayer strongly deviates from the 
hydrogenic Rydberg series \cite{Chernikov_2014,Qiu_2013}, which indicates a significantly 
reduced and non-local dielectric screening of the Coulomb interaction. This is further supported 
by the fact that monolayers of $1T$-NbSe$_2$ and $1T$-TaS$_2$ are Mott insulators, which is 
very rare for $4d$ and $5d$ transition metal compounds \cite{Liu_2021}. For metallic 2D TMDs, 
localized plasmons have been observed experimentally in the correlated 2D CDW $1H$-TaSe$_2$ and 
$1H$-NbSe$_2$ materials  \cite{Song_2021,Cudazzo_2016,da2020universal,da_Jornada_2020}. 
Additionally, electron energy-loss spectra (EELS) for these materials have revealed a negative 
dispersion of the plasmons in these materials \cite{van_Wezel_2011}, which is in contrast to 
the results for a homogeneous 2D electron gas. This unconventional behavior of the plasmon
dispersion is attributed to the electronic correlation effects in these materials.

Monolayer $1T'$ TMDs, especially those with heavy elements like W and Mo have strong spin-orbit 
coupling and are topological insulators with one-way linear dispersion bands near the Fermi level E$_F$ 
\cite{Chen_2018,Ugeda_2018}. This crystal phase exhibits an intriguing screening behavior, 
as electrons in linear bands do not effectively screen long-range Coulomb 
interactions \cite{Wehling_2011,Hwang_2007,Hadipour_2018,_a_o_lu_2017}. This property provides a 
unique perspective on the interplay between electronic structure and screening phenomena in 
these materials. Additionally, the topological insulating behavior of $1T'$ TMDs adds another layer 
of complexity to the interplay of electronic properties and their implications for device applications 
and emergent physical phenomena.

The study of electronic screening effects and the calculation of Coulomb matrix elements in 
transition metal (TM) materials, including TMDs, have been the subject of several studies 
\cite{TM_cRPA,Karbalaee_Aghaee_2022,yekta_2021,Hadipour_2019}. Most of these studies have 
focused on $3d$ TM compounds, which have narrow $t_{2g}$ and $e_g$ states and strong correlation 
effects. A few works have investigated Coulomb interactions in TMDs with $4d$ elements 
\cite{Kim_2023,Darancet_2014,van2018competing}. For example, in the distorted structure of 
TaS$_2$, the on-site Coulomb interaction $U_{00}$ was calculated to be 0.65 eV \cite{Kim_2023},
which is significantly lower than the value of 2.27 eV for undistorted $1T$-TaS$_2$ \cite{Darancet_2014}. 
Van Loon et al. used first-principles calculations to determine an effective on-site Coulomb 
interaction $U_{00}=1.8$\,eV and a nearest-neighbor interaction $U_{01}=1$\,eV for $1H$-NbS$_2$ \cite{van2018competing}. These effective interactions can be incorporated into model Hamiltonians, 
which can improve the predictive power of model-based calculations. This motivates our systematic
and fully \emph{ab-initio} approach to comprehensively compute effective Coulomb parameters for $4d$ and $5d$
TMD monolayers. We anticipate that our work will provide new insights into the complex interplay 
between electronic structure and interaction effects and that it will enhance the predictive 
power and applicability of theoretical models in this field.

In this study, we used first-principles electronic structure calculations and constrained 
random-phase approximation (cRPA) within the full-potential linearized augmented-plane-wave 
(FLAPW) method to calculate the effective Coulomb interaction parameters between localized 
$d$ electrons in 2D TMDs. We specifically explored materials represented by the formula MX$_2$ 
(M=Nb, Ta, Mo, W, and X=S, Se, Te) and considered three different phases ($1H$, $1T$, and $1T'$).
All compounds in the $1T$ and $1T'$ phases are metallic, while Mo and W-based compounds in the 
$1H$ phases are semiconductors and Nb and Ta-based compounds in the same phase exhibit cold 
metallic behavior (see Fig.\,\ref{fig0}  and Table~\ref{table:lattice_parameter}). Our results show that the short-range 
interactions  are strongly screened in all three phases, whereas the long-range
interaction remains significant even in metallic systems. This nonconventional screening 
provides a compelling explanation for the deviations observed in the usual hydrogenic Rydberg 
series and conventional plasmon dispersion in 2D semiconducting and metallic TMDs, respectively.
Our calculations yield on-site Coulomb interaction values within the ranges of 0.8-2.5 eV, 0.8-1.9 eV, 
and 0.9-2.4 eV for the $1H$, $1T$, and $1T'$ structures, respectively. We find that these values depend 
on the specific chalcogen X, the number of $d$ electrons, and the correlated subspace. 
Using the calculated $U$ parameters for undistorted $1T$ structure, we extract the on-site 
effective $U^{\mathrm{eff}}_{00}$ and nearest neighbor $U^{\mathrm{eff}}_{01}$ Coulomb 
interaction parameters for reconstructed  commensurate CDW NbX$_2$ and TaX$_2$ compounds. Furthermore, 
our findings indicate a substantially high ratio of on-site Coulomb interaction to bandwidth 
($U^{\mathrm{eff}}_{00}/W_b \gg 1$) in CDW TMDs, providing robust evidence for the experimentally observed strongly 
correlated Mott phase. The rest of the manuscript is organized as follows: In Sec.\,II, we 
outline the computational method. Sec.\,III covers results and discussions. Finally, Sec.\,IV 
presents conclusions and a summary.

\section{Computational Method}{\label{computational metod}}

\subsection{Crystal structure and ground state calculations}{\label{sec:crystal}}

\begin{figure*}
\centering
\includegraphics[width=1.0\linewidth]{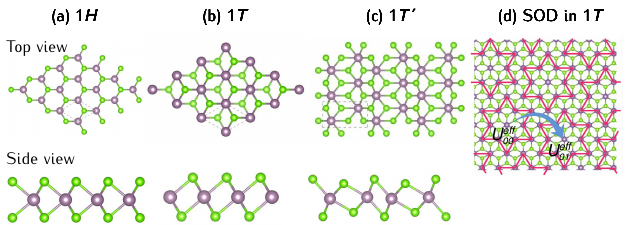}
\caption{Side and top views of the two-dimensional crystal structure of transition 
metal dichalcogenides MX$_2$ in (a) 1$H$ structure, (b) 1$T$ structure, (c) 1$T^\prime$ structure, 
and (d) Star of David (SOD) reconstructed 1$T$ crystal structure. The purple and green 
spheres exhibit M and X atoms, respectively.}
\label{fig:struct}
\end{figure*}

We consider 2D TMDs, which have the chemical formula MX$_2$. Here, M represents elements such 
as Mo, W, Nb, and Ta, and X represents chalcogen elements, namely S, Se, and Te. Our study 
encompasses TMDs with distinct crystallographic structures, including trigonal prismatic 
(1$H$), octahedral (1$T$), and distorted octahedral (1$T'$) structures. The crystal structures 
are shown in Fig.\,\ref{fig:struct} and the corresponding lattice parameters are given in Table~\ref{table:lattice_parameter}. 
In the 1$H$ and 1$T$ structures, the fundamental unit cell has a hexagonal lattice configuration
and it contains one M atom and two X atoms, which are separated by a vacuum region of 20\,\AA. This is 
illustrated in Fig.\,\ref{fig:struct}(a) and \,\ref{fig:struct}(b). For our investigation of the $1T'$ structure, we use an 
orthorhombic unit cell containing two M atoms and four X atoms, as shown in Fig.\,\ref{fig:struct}(c).

\begin{table}
\caption{Lattice parameters and corresponding ground states of the 
studied MX$_2$ (M=Mo, W, Nb, Ta and X=S, Se, Ta) compounds 
in 1$H$, 1$T$, and 1$T^{\prime}$ structures. Lattice parameters are taken from Ref.\,\cite{haastrup2018computational}. }
\label{table:lattice_parameter}
\centering
\begin{ruledtabular}
\begin{tabular}{lclll}
MX$_2$    & Phase    & \emph{a} (\AA) & \emph{b} (\AA) & Ground state  \\ \hline
MoS$_2$   & ${1H}$          & 3.18 & 3.18 & Semiconductor \\				
          & ${1T}$          & 3.19 & 3.19 & Metal         \\
	   & ${1T^{\prime}}$ & 5.72 & 3.18 & Semimetal     \\  
MoSe$_2$  &${1H}$           & 3.32 & 3.32 & Semiconductor  \\
          &${1T}$           & 3.28 & 3.28 &  Metal         \\
	   & ${1T^{\prime}}$ & 5.96 & 3.29 & Semimetal      \\ 	
MoTe$_2$ & ${1H}$           & 3.55 & 3.55 & Semiconductor  \\
         & ${1T}$           & 3.49 & 3.49 & Metal          \\
         & ${1T^{\prime}}$  & 6.36 & 3.46 & Metal          \\ 	
WS$_2$   & ${1H}$           & 3.19 & 3.19 & Semiconductor  \\
         & ${1T}$           & 3.21 & 3.21 & Metal          \\
         & ${1T^{\prime}}$  & 5.73 & 3.20 & Semimetal      \\ 	
WSe$_2$  & ${1H}$           & 3.32 & 3.32 & Semiconductor  \\
         & ${1T}$           & 3.29 & 3.29 & Metal          \\
         & ${1T^{\prime}}$  & 5.95 & 3.30 & Semimetal      \\ 	
WTe$_2$  & ${1H}$           & 3.55 & 3.55  & Semiconductor \\
         & ${1T}$           & 3.51 & 3.51  & Metal         \\
         & ${1T^{\prime}}$  & 6.31 & 3.49  & Metal         \\  
NbS$_2$  & ${1H}$ & 3.34 &  3.34 & Cold-metal  \\
         & ${1T}$ & 3.38 &  3.38 & Metal        \\  
NbSe$_2$ & ${1H}$   &  3.47 &  3.47  & Cold-metal  \\
         & ${1T}$   & 3.48  &  3.48  &  Metal      \\  
NbTe$_2$ & ${1H}$   & 3.68 & 3.68 & Cold-metal    \\
         & ${1T}$   & 3.65 & 3.65 & Metal         \\  
TaS$_2$  & ${1H}$  & 3.34 & 3.34  & Cold-metal    \\
         & ${1T}$  & 3.38 & 3.38  & Metal         \\ 	
TaSe$_2$ & ${1H}$  & 3.47 & 3.47  & Cold-metal  \\
         & ${1T}$  & 3.50 & 3.50  & Metal       \\	
TaTe$_2$ & ${1H}$  & 3.70 & 3.70  & Cold-metal    \\
         & ${1T}$  & 3.69 & 3.69  & Metal \\		
\end{tabular}
\end{ruledtabular}
\end{table}

The variation in crystal field splitting induced by the neighboring chalcogen X atoms in 
the three lattice configurations of 1$H$-MX$_2$, 1$T$-MX$_2$, and 1$T'$-MX$_2$ yields distinct correlated 
subspaces, which will be discussed in the following section. This phenomenon plays a key role in expressing the differences in the observed 
electronic and optical properties of the materials.

In the $1H$ structure, the chalcogens are aligned vertically along the $z$-axis, with the 
transition metals sandwiched within the central plane along the $x$-axis. This arrangement 
results in a Bernal (ABA) stacking configuration, as depicted in the side view of 
Fig.\,\ref{fig:struct}(a). Conversely, the $1T$ structure exhibits rhombohedral (ABC) stacking, 
as evident from the side view presented in Fig.\,\ref{fig:struct}(b)\cite{manzeli20172d,wang2018crystal}. 
In the $1H$ structure, the $d$ orbitals experience a splitting into a singlet $d_{z^2}$, an 
intermediate-energy doublet  $e_g$ ($d_{x^2-y^2}$, $d_{xy}$), and a high-energy doublet 
$e^\prime_g$ ($d_{yz}$, $d_{xz}$). On the other hand, in the case of the 1$T$ structure, 
the $d$ orbital splits into three lower-energy $t_{2g}$ ($d_{xy}$, $d_{yz}$, $d_{xz}$) 
and two higher-energy $e_g$ ($d_{x^2-y^2}$, $d_{z^2}$) states \cite{Qian_2020}.

The 1$T'$-MX$_2$ structure of TMDs is a low-symmetry crystal phase that can be considered 
as a periodically distorted structure of the 1$T$ structure. In the 1$T'$ structure, 
the two adjacent transition metal (TM) atoms move towards each other in the $y$-direction, 
compared to the 1$T$ structure. The 1$T^{\prime}$ structure is more stable than the 1$T$ 
structure, and the energy barrier to separate the 1$T$ structure from the stable 
1$T'$ structure is nearly zero, leading to spontaneous structural distortions \cite{chen2020charge,voiry2015phase}. 
This distortion causes an inversion of the band structure at the  $\Gamma$ point between 
the $p_x$ orbital of the dichalcogenide and the $d$ orbital of the TM, and a cone-like band 
structure is formed \cite{choe2016understanding,zhao2018metastable}.

The FLAPW method, as implemented in the FLEUR code \cite{fleurWeb}, is used for the 
ground-state calculations. We employ the generalized gradient approximation (GGA) 
to the exchange-correlation potential, as parametrized by Perdew et al. \cite{GGA_1}. 
To ensure that the results are consistent, calculations are performed using the same 
cutoff for the wave functions (k$_{\mathrm{max}}$ = 4 a.u.$^{-1}$), and the same 
16 $\times$ 16 $\times$ 1 (12 $\times$ 16 $\times$ 1)  $\mathbf{k}$-point grid for 
the 1$H$ and 1$T$ (1$T'$) structures in the determination of the ground states. 
These parameters have been verified to yield well-converged Coulomb interaction 
parameters across all studied compounds. The maximally localized Wannier functions  
(MLWFs) are constructed using the WANNIER90 library \cite{pizzi2020wannier90,Wannier90,Wannier901,Fleur_Wannier90}. 
The effective Coulomb potential is calculated within the constrained random-phase 
approximation (cRPA) method \cite{aryasetiawan2004frequency,aryasetiawan2006calculations,miyake2009ab,nomura2012effective,shih2012screened}, 
as implemented in the SPEX code \cite{schindlmayr2010first,friedrich2010efficient}. We use 
a 14$\times$14$\times$1 (10$\times$14$\times$1) $\mathbf{k}$-point grid for 1$H$ and 1$T$ (1$T'$) 
structures in the cRPA calculations.

\subsection{The cRPA method and parametrization of the Coulomb matrix}{\label{sec:cRPA}}

The cRPA method is an efficient way to calculate the screened Coulomb interaction between 
localized electrons. It allows us to determine individual Coulomb matrix elements, such 
as on-site, off-site, intra-orbital, inter-orbital, and exchange, as well as their 
frequency dependence. The fully screened Coulomb interaction $W$ is related to the bare
Coulomb interaction $V$ by	
\begin{equation}\label{eq.1}
W(\mathbf{r},\mathbf{r^\prime};\omega) =
\int \mathrm{d}\mathbf{r^{\prime\prime}}
\epsilon^{-1}(\mathbf{r},\mathbf{r^{\prime\prime}},\omega) 
V(\mathbf{r^{\prime\prime}},\mathbf{r^\prime}) ,
\end{equation}
where $	\epsilon(\mathbf{r},\mathbf{r^{\prime\prime}},\omega) $ is the dielectric function. 
The dielectric function is related to the electron polarizability $P$ by
\begin{equation}\label{eq.2}
\epsilon(\mathbf{r},\mathbf{r^{\prime}},\omega) =
\delta (\mathbf{r}-\mathbf{r^{\prime}})
- \int \mathrm{d}\mathbf{r^{\prime\prime}}
V (\mathbf{r},\mathbf{r^{\prime\prime}})
P (\mathbf{r^{\prime\prime}},\mathbf{r^{\prime}},\omega) ,
\end{equation}
where the RPA polarization function is given by
\begin{align}\label{eq.3}
P\left(\mathbf{r},\mathbf{r^\prime};\omega\right) &= 2 \sum_{m}^{\mathrm{occ}}{\sum_{m^\prime}^{\mathrm{unocc}}}{\varphi_{m}(\mathbf{r})\varphi_{m^\prime}^{*}(\mathbf{r})\varphi_{m}^{*}(\mathbf{r^\prime})
			\varphi_{m^\prime}(\mathbf{r^\prime})
		}
		\nonumber \\
		&
		\times 
		\bigg[
		\frac{1}{\omega-\Delta_{mm^\prime} + i\eta}
		-
		\frac{1}{\omega-\Delta_{mm^\prime} - i\eta}
		\bigg] ,
	\end{align}
Here, $\varphi_{m} (\mathbf{r})$ are the single-particle DFT Kohn-Sham eigenfunctions, and $\eta$ a positive infinitesimal. $\Delta_{mm^\prime}=\epsilon_{m^\prime}-\epsilon_{m}$
with the Kohn-Sham eigenvalues $\epsilon_{m}$. 
In the cRPA approach, to exclude the screening due to the correlated subspace, 
we partition the full polarization function of Eq.\,(\ref{eq.3}) into two parts.
\begin{equation}\label{eq.4}
P = P_d + P_r ,
\end{equation}
where $P_d$ includes only the transitions $(m \rightarrow m^{\prime})$ between the
states of the correlated subspace and  $P_r$ is the remainder.
Then, the frequency-dependent effective Coulomb interaction
is given schematically by the matrix equation
\begin{equation}\label{eq.5}
U(\omega)=[1-VP_r(\omega)]^{-1}V ,
\end{equation}
The set $P_r$ comprises all transitions, excluding those occurring within the correlated subspace
The matrix elements of the effective Coulomb interaction in the MLWF basis is given by
\begin{align}\label{eq.6}
{U}_{\mathbf{R}n_1,n_2,n_3,n_4}(\omega) = & 
\int\int
\mathrm{d}\mathbf{r} 
\mathrm{d}\mathbf{r^\prime} w_{n_1\mathbf{R}}^*(\mathbf{r})
w_{n_3\mathbf{R}}(\mathbf{r})
U(\mathbf{r},\mathbf{r^\prime},\omega)
\nonumber \\ &  
\times
w_{n_4\mathbf{R}}^*(\mathbf{r^\prime})
w_{n_2\mathbf{R}}(\mathbf{r^\prime}) ,
\end{align}
where $w_{n\mathbf{R}} (\mathbf{r})$ is the MLWF at site $\mathbf{R}$ with orbital 
index $n$, and the effective Coulomb potential $U(\mathbf{r},\mathbf{r^\prime},\omega)$ 
is calculated within the cRPA method as described above. We define the average on-site 
diagonal (direct intra-orbital) $U$ and off-diagonal (direct and exchange 
inter-orbital) $U^\prime$, $J$ matrix elements of the screened Coulomb potential 
in the static limit $(\omega =
0)$ as follows \cite{anisimov1993density,anisimov2010electronic}:
\begin{align}\label{eq.7}
	{U}=\frac{1}{L}\sum_{m}U_{mm;mm} ,
\end{align}
\begin{align}\label{eq.8}
	{U^\prime}=\frac{1}{L(L-1)}\sum_{m\neq n}U_{mn;mn} ,
\end{align}
\begin{align}\label{eq.9}
	J=\frac{1}{L(L-1)}\sum_{m\neq n}U_{mn;nm} ,
\end{align}
where $L$ is the number of localized orbitals, i.e., one for $d_{z^2}$, three 
for $d_{z^2},d_{xy},d_{x^2-y^2}$ and five for full $d$ orbitals. 
We employ Eq.\,(\ref{eq.7}) to Eq.\,(\ref{eq.9}) for all subspaces discussed in this paper.
 One can show that these Hubbard-Kanamori parameters define the full partially screened Coulomb matrix [Eq.\ (\ref{eq.6})] of subspaces formed by $t_{2g}$ and $e_g$  orbitals (assuming spherical symmetry around the atoms).
However, to fully define Eq.\ (\ref{eq.6}) for the 
three-orbital 
(${d_{z^2}+d_{xy}+d_{x^2\mbox{-}y^2}}$) and five-orbital ($d$) subspaces, we need at least one additional matrix element. For reference, we provide explicit values for several relevant matrix elements in the supplementary material.
Similar to the definition of $U$ ($U^\prime$, $J$), we can 
also define the so-called fully screened interaction $W$ as well as unscreened (bare) $V$. 
The bare Coulomb interaction parameter $V$ provides information on the localization of Wannier functions.
Several procedures have been proposed in the literature to calculate the polarization
function for entangled bands \cite{TM_cRPA,csacsiouglu2012strength,aryasetiawan2004frequency,aryasetiawan2006calculations,miyake2009ab,nomura2012effective,shih2012screened}. 
In the present work, we use the method described in Ref.\,\cite{TM_cRPA,csacsiouglu2012strength}.

Calculating Coulomb interaction parameters for the $1H$, $1T$, and $1T'$ structures 
is relatively straightforward. 
However, the star-of-David (SOD) reconstruction increases the number of atoms per 2D unit cell to 39, of which 13 are TM atoms. 
cRPA calculations for such systems would become very demanding. 
For the sake of simplicity, we therefore utilize the results from the 
undistorted $1T$ structure to make estimations of the on-site and long-range 
effective Coulomb interactions \cite{Darancet_2014,Kamil_2018,Kim_2023} for the SOD reconstructed systems. The estimated effective 
Coulomb interaction is given by \cite{Kim_2023}
\begin{equation}\label{eq.10}
U^{\mathrm{eff}} = \frac{1}{13^2}\sum_{\mathbf{R},\mathbf{R^\prime}} U_{\mathbf{R}-\mathbf{R^\prime}},
\end{equation}
when both the vectors $\mathbf{R}$ and $\mathbf{R^\prime}$ correspond to the positions of 
TM atoms within a specific star in the 1$T$ structure, $U^{\mathrm{eff}}$ represents the on-site 
effective interaction $U^{\mathrm{eff}}_{00}$ for that star. On the other hand, to determine 
the long-range interaction $U^{\mathrm{eff}}_{01}$ within the reconstructed lattice, vector 
$\mathbf{R}$ should pertain to star A, while $\mathbf{R^\prime}$ refers to the indices of TM atoms 
belonging to the nearest neighboring star B [see a schematic representation of the star of 
David in Fig.\,\ref{fig:struct}(d)].

\section{\label{sec:result}Results and discussion}
 
\subsection{Correlated subspace}{\label{sec:subspace}}  

\begin{figure}
\centering
\includegraphics[width=1.0\linewidth]{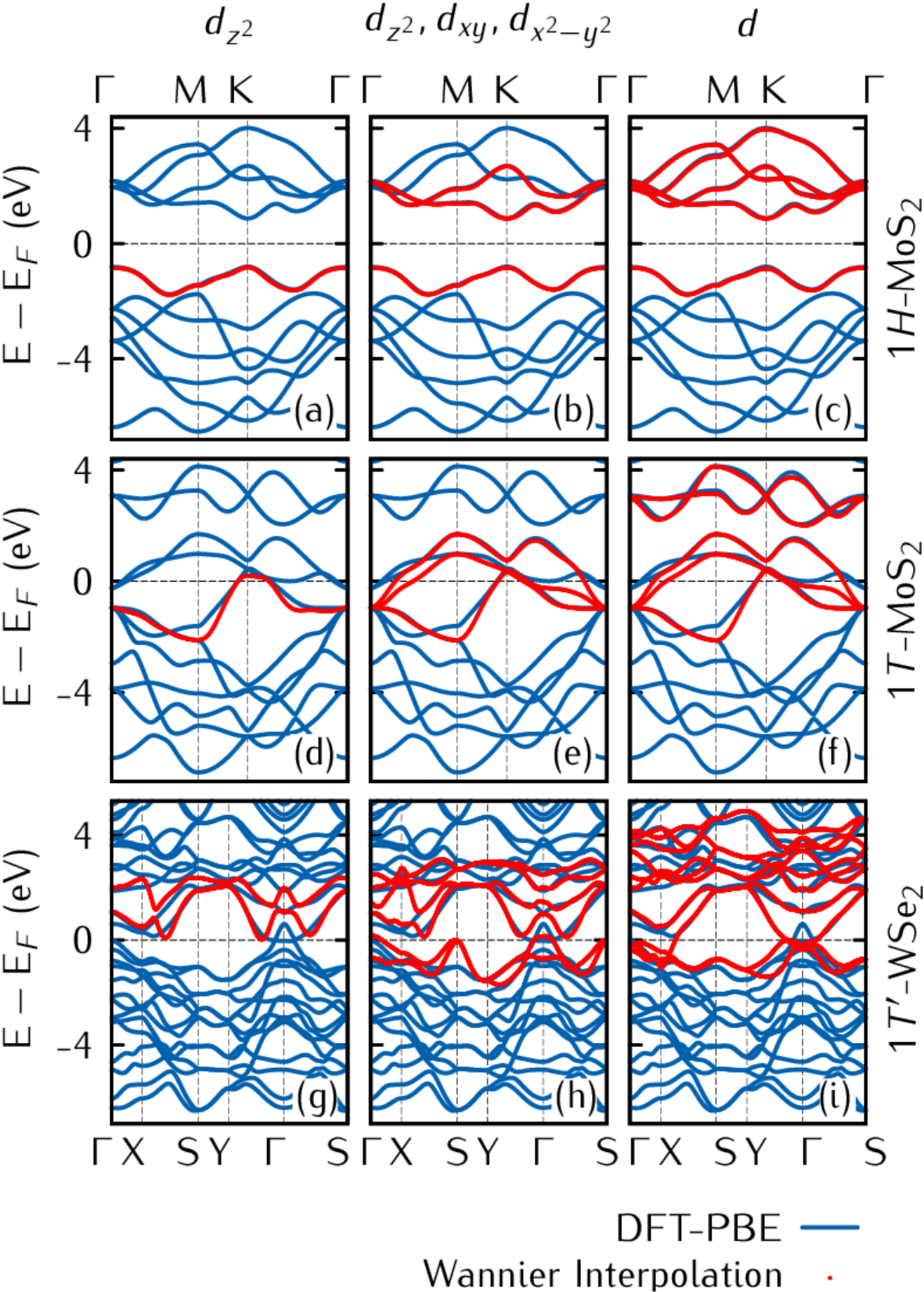}
\caption{(Color online) DFT-PBE (blue) and Wannier-interpolated band structures
(red) of (a) (b) (c) 1$H$-MoS$_2$ and (d) (e) (f) 1$T$-MoS$_2$, and (g) (h) (i) 
1$T^{\prime}$-WSe$_2$.  In each system, we considered three correlated subspaces derived 
from one-orbital $d_{z^2}$, three-orbital $d_{z^2},d_{xy},d_{x^2-y^2}$, and full $d$ orbitals of the TM atom.}
\label{fig:interpol-MoW}
\end{figure}

\begin{figure}
\centering
\includegraphics[width=1.0\linewidth]{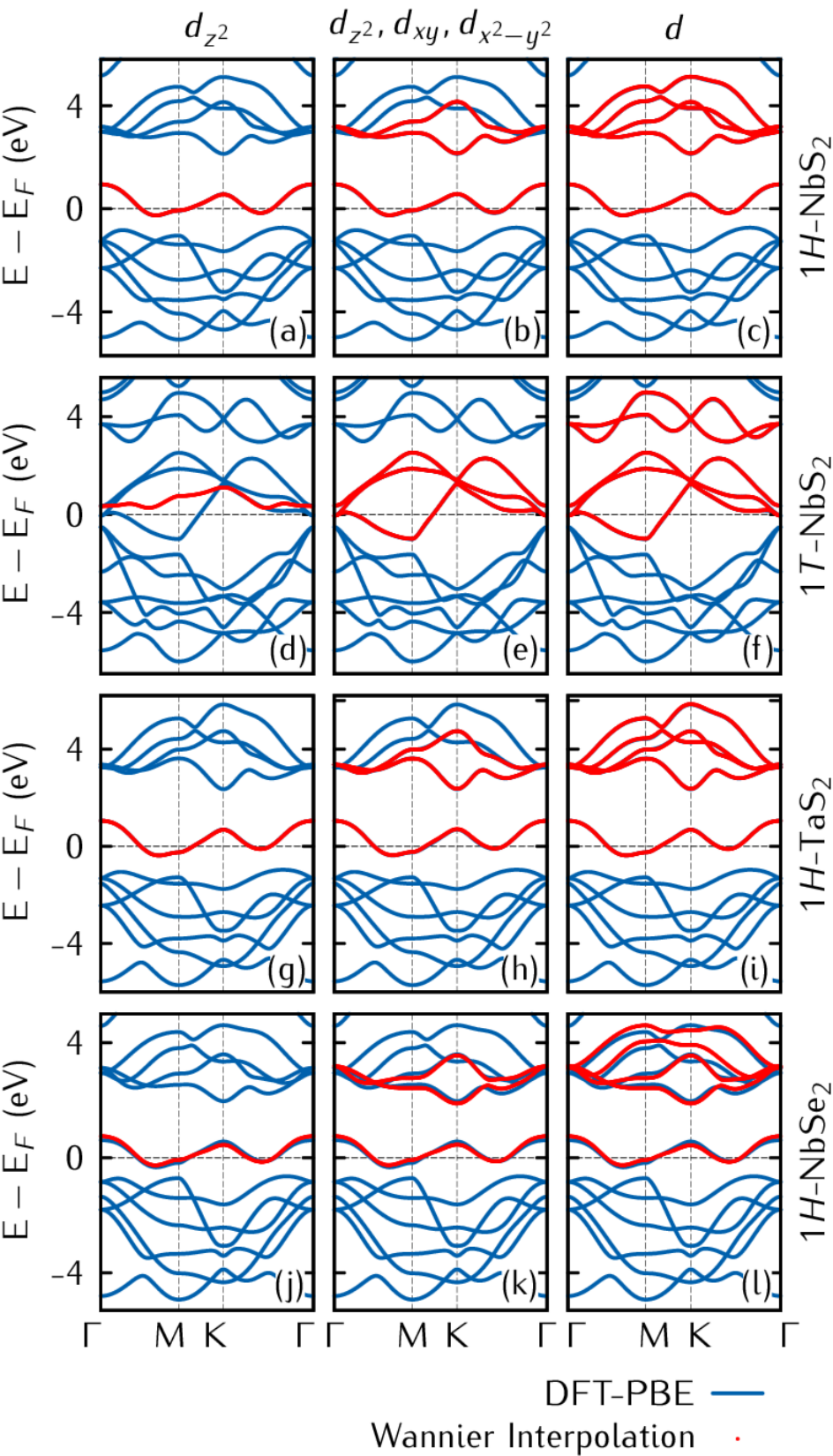}
\caption{(Color online)  DFT-PBE (blue) and Wannier-interpolated band structures
(red) of (a) (b) (c) 1$H$-NbS$_2$ and (d) (e) (f) 1$T$-NbS$_2$, (g) (h) (i) 1$H$-TaS$_2$, and (j) 
(k) (l) 1$H$-NbSe$_2$. In each system, we considered three correlated subspaces derived 
from one-orbital $d_{z^2}$, three-orbital $d_{z^2},d_{xy},d_{x^2-y^2}$, and full $d$ orbitals of the  TM atom.}
\label{fig:interpol-NbTa}
\end{figure}

\begin{figure}
\centering
\includegraphics[width=1.0\linewidth]{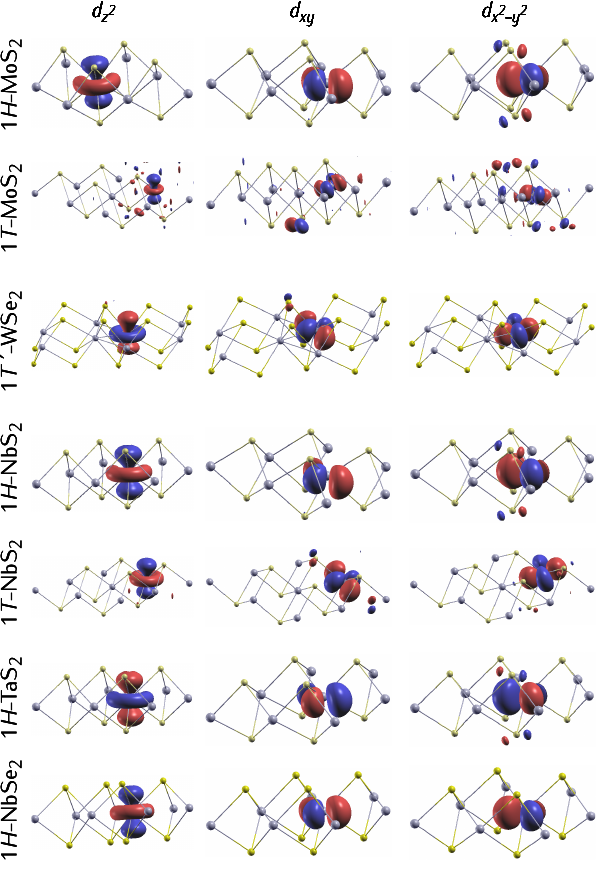}
\caption{Plot of MLWFs for TM atoms in 1$H$-MoS$_2$, 1$T$-MoS$_2$, 1$T^{\prime}$-WSe$_2$, 1$H$-NbS$_2$, 1$T$-NbS$_2$, 1$H$-TaS$_2$, and 1$H$-NbSe$_2$. First column: The $d_{z^2}$-like
MLWFs, considering only one-orbital subspace, i-e., $d_{z^2}$. Second column: The $d_{xy}$-like MLWFs, considering a three-orbital ($d_{z^2}+ d_{xy} + d_{x^2\mbox{-}y^2}$) subspace. Third
column: The $d_{x^2\mbox{-}y^2}$like MLWFs, considering the full five orbital $d$-space.
Same isovalue is used in all cases.}
\label{fig:Wannier_orbitals}
\end{figure}

In order to determine the strength of the screened Coulomb interaction, it is important 
to identify the correlated subspace. This is the subspace of electronic states that are 
most strongly interacting, and it is essential for the accurate construction 
of Wannier functions and the corresponding effective low-energy model Hamiltonian.
As a first step, we performed electronic structure calculations for all systems. 
The projected band structures (see Figs.\,S1 and S2 in supplementary material) show that the $d$ orbitals 
of the TM atom make a significant contribution to the bands near the Fermi level, compared to 
the other orbitals from chalcogen atoms. Thus, all investigated compounds can be described 
by an effective low-energy model based on only TM atom $d$ electrons. In Fig.\,\ref{fig:interpol-MoW} 
and Fig.\,\ref{fig:interpol-NbTa} we compare the DFT-PBE band structures with the corresponding 
Wannier-interpolated bands for some selected materials. The corresponding Wannier orbitals 
are depicted in Fig.\,\ref{fig:Wannier_orbitals}. As seen in Fig.\,\ref{fig:interpol-MoW}, the 1$H$ phase
of the  MX$_2$ (M=Mo, W; X=S, Se, Te) compounds can be described  well by a three-orbital
($d_{z^{2}}+e_{g}$) model, while the  1$T$ and 1$T'$  phases of the same materials
require a full $d$-orbital effective model. On the other hand, 1$H$ MX$_2$ (M=Nb, 
Ta; X=S, Se, Te) compounds show cold metallic behavior and thus they can be described by a 
simple one-orbital model, while the same materials in the 1$T$ structure require a 
three-orbital model. For consistency, the  Coulomb interaction parameters in the following
section will be presented for a one-orbital ($d_{z^2}$), for a three-orbital
($d_{z^2}+ d_{xy} +d_{x^2-y^2}$), and for five-orbital ($d$) correlated subspaces.

\subsection{Coulomb interaction parameters: MX$_2$ (M=Mo, W; X=S, Se, Te)}
\label{}

Prior to discussing the effective Coulomb interaction parameters in
in MX$_2$ (M=Mo, W; X=S, Se, Te) compounds in 1$H$, 1$T$, and 1$T'$
structures it is  worth noting that the screening of the Coulomb interaction
in 2D semiconductors has been extensively explored in recent years by numerous 
researchers employing various methodologies. For instance, the 
quantum-electrostatic heterostructure model 
\cite{Andersen_2015} employs a monopole-dipole approximation to estimate the 
dielectric function at zero frequency. 
Trolle et al.\ \cite{Trolle_2017} proposed a model 2D dielectric function to determine excitonic binding energies.
One difficulty in the calculation of dielectric functions of 2D systems with a 3D code is the unwanted but unavoidable screening contribution of the periodic images of the layer in the neighboring supercells. 
A possible solution is to truncate the Coulomb interaction in the $z$ direction \cite{H_ser_2013}. As an alternative, we employ an extrapolation formula that yields the dielectric function for infinite inter-layer distances \cite{Nazarov14,Rost23}. The effect of the extrapolation can be seen in Fig.\ \ref{fig:extrapolate}.
In Fig.\,S5 of the supplementary material, we show, for reference, the $q$-dependent static 
dielectric function $\epsilon(\omega=0,q)$ calculated within RPA for 1H-MoS$_2$ and a layer distance (supercell height) of 25\AA, thus \emph{including} the inter-layer screening.
The calculated dielectric function exhibits a behavior similar to the results obtained in other works \cite{H_ser_2013,Molina_S_nchez_2011,Ramasubramaniam_2012,Cheiwchanchamnangij_2012}.


\begin{figure}
\centering
\includegraphics[width=1.0\linewidth]{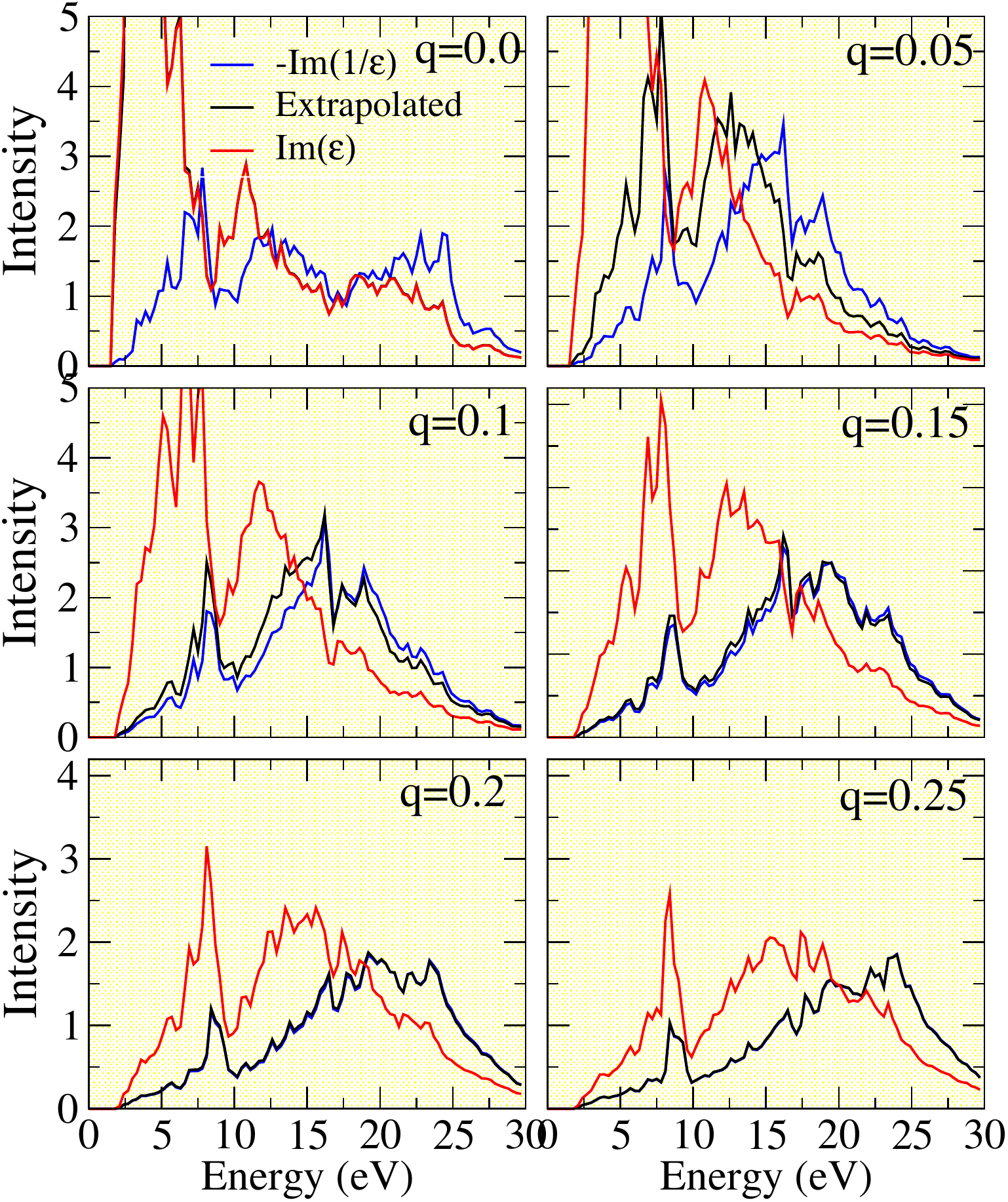}
\caption{(Color online) Loss function for 1$H$-MoS$_2$. The extrapolated curve corresponds to a monolayer (infinite layer distance) and tends towards $-\text{Im}(1/\varepsilon)$ and $\text{Im}(\varepsilon)$ for large and small $q$, respectively.}
\label{fig:extrapolate}
\end{figure}

The choice of correlated subspace can significantly 
impact the accuracy of the calculated properties. For example, a minimal three-orbital 
low-energy model might be sufficient for investigating transport properties in 1$H$ structure, 
but the inclusion of the full $d$-orbital correlated subspace might be necessary to accurately 
account for the complex interplay of electrons and photons within the material when delving 
into optical properties, such as absorption and emission spectra. This distinction highlights 
the nuanced nature of these materials' behaviors and the necessity of tailoring the correlated 
subspace according to the specific properties under investigation. As mentioned in the preceding 
section, we will consider a one-orbital (${d_{z^2}}$), a three-orbital 
(${d_{z^2}+d_{xy}+d_{x^2\mbox{-}y^2}}$), and five-orbital ($d$) correlated subspace.

\begin{table*}
\caption{Bare Coulomb interaction $V$, partially screened Hubbard-Kanamori parameters 
[$U$,  $U^{\prime}$, $J$ (in eV)] and fully screened $W$ (in eV) for $d$ orbitals of 
TMs in MX$_2$ (M=Mo, W; X=S, Se, Te) compounds.}
\centering
\begin{ruledtabular}
\begin{tabular}{cccccccc}
MX$_2$  &Phase &  Orbitals & $V^{}$(eV)&$U^{}$(eV)& ${U^{\prime}}(eV)^{}$  & $J(eV)^{}$ & $W^{}$(eV)  \\ \hline	
$ \rm{  } $ & ${1H}$ & ${d_{z^2}}$ & {8.84} & {1.95}  & {-----} & {-----} & {1.95}  \\	
&  & ${d_{z^2},d_{xy},d_{x^2-y^2}}$ & {9.47} & {2.33}  & {2.04} & {0.15} & {1.54}  \\
&  & ${d}$ & {9.08} & {2.30}  & {2.03} & {0.13} & {1.54}  \\ \addlinespace  $ \rm{MoS_2} $ & ${1T}$ & ${d_{z^2}}$ & {9.53} & {0.45}  & {-----} & {-----} & {0.44}  \\	
&  & ${d_{z^2},d_{xy},d_{x^2-y^2}}$ & {11.32} & {1.28}  & {0.96} & {0.17} & {0.64}  \\
&  & ${d}$ & {10.39} & {1.27}  & {1.01} & {0.13} & {0.60}  \\ \addlinespace  $ \rm{  } $ & $ \mathrm{1T^{\prime}}$ & $ {d_{z^2}}$ & {8.37} & {1.66}  & {-----} & {-----} & {1.54}  \\	
&  & $ {d_{z^2},d_{xy},d_{x^2-y^2}}$ & {8.93} & {2.00}  & {1.57} & {0.22} & {1.41}  \\
&  & $ {d}$ & {8.90} & {2.18}  & {1.77} & {0.21} & {1.45}  \\ \hline 
$ \rm{  } $ & $ \mathrm{1H}$ & $ {d_{z^2}}$ & {8.91} & {1.96}  & {-----} & {-----} & {1.96}  \\	
&  & $ {d_{z^2},d_{xy},d_{x^2-y^2}}$ & {9.36} & {2.26}  & {1.98} & {0.16} & {1.49}  \\
&  & $ {d}$ & {9.16} & {2.26}  & {1.99} & {0.13} & {1.51}  \\ \addlinespace  $ \rm{MoSe_2} $ & $ {1T}$ & $ {d_{z^2}}$ & {8.94} & {0.49}  & {-----} & {-----} & {0.49}  \\	
&  & $ {d_{z^2},d_{xy},d_{x^2-y^2}}$ & {10.55} & {1.41}  & {1.06} & {0.18} & {0.65}  \\
&  & $ {d}$ & {9.59} & {1.13}  & {0.89} & {0.11} & {0.60}  \\ \addlinespace  $ \rm{  } $ & $ {1T^{\prime}}$ & $ {d_{z^2}}$ & {8.36} & {2.30}  & {-----} & {-----} & {2.30}  \\	
&  & $ {d_{z^2},d_{xy},d_{x^2-y^2}}$ & {9.08} & {1.89}  & {1.43} & {0.21} & {1.46}  \\
&  & $ {d}$ & {9.57} & {2.28}  & {1.77} & {0.23} & {1.57}  \\ \hline 
$ \rm{  } $ & $ {1H}$ & $ {d_{z^2}}$ & {6.12} & {1.08}  & {-----} & {-----} & {1.08}  \\	
&  & $ {d_{z^2},d_{xy},d_{x^2-y^2}}$ & {6.81} & {1.35}  & {1.18} & {0.08} & {0.95}  \\
&  & $ {d}$ & {7.28} & {1.46}  & {1.25} & {0.09} & {1.04}  \\ \addlinespace  $ \rm{MoTe_2} $ & $ {1T}$ & $ {d_{z^2}}$ & {6.69} & {0.49}  & {-----} & {-----} & {0.49}  \\	
&  & $ {d_{z^2},d_{xy},d_{x^2-y^2}}$ & {7.03} & {0.52}  & {0.35} & {0.07} & {0.60}  \\
&  & $ {d}$ & {8.14} & {1.53}  & {1.31} & {0.11} & {0.60}  \\ \addlinespace  $ \rm{  } $ & $ {1T^{\prime}}$ & $ {d_{z^2}}$ & {5.97} & {0.55}  & {-----} & {-----} & {0.55}  \\	
&  & $ {d_{z^2},d_{xy},d_{x^2-y^2}}$ & {6.81} & {0.85}  & {0.56} & {0.15} & {0.65}  \\
&  & $ {d}$ & {7.96} & {1.14}  & {0.78} & {0.18} & {0.81}  \\ \hline 
$ \rm{  } $ & $ {1H}$ & $ {d_{z^2}}$ & {8.43} & {1.98}  & {-----} & {-----} & {1.98}  \\	
&  & $ {d_{z^2},d_{xy},d_{x^2-y^2}}$ & {8.85} & {2.42}  & {2.16} & {0.15} & {1.62}  \\
&  & $ {d}$ & {8.47} & {2.39}  & {2.14} & {0.12} & {1.62}  \\ \addlinespace  $ \rm{WS_2} $ & $ {1T}$ & $ {d_{z^2}}$ & {8.45} & {0.40}  & {-----} & {-----} & {0.40}  \\	
&  & $ {d_{z^2},d_{xy},d_{x^2-y^2}}$ & {9.56} & {1.09}  & {0.85} & {0.13} & {0.61}  \\
&  & $ {d}$ & {8.85} & {1.31}  & {1.09} & {0.12} & {0.58}  \\ \addlinespace  $ \rm{  } $ & $ {1T^{\prime}}$ & $ {d_{z^2}}$ & {7.94} & {1.40}  & {-----} & {-----} & {1.40}  \\	
&  & $ {d_{z^2},d_{xy},d_{x^2-y^2}}$ & {8.06} & {1.90}  & {1.51} & {0.21} & {1.33}  \\
&  & $ {d}$ & {8.13} & {2.08}  & {1.70} & {0.20} & {1.38}  \\ \hline 
$ \rm{  } $ & $ {1H}$ & $ {d_{z^2}}$ & {8.31} & {2.04}  & {-----} & {-----} & {2.04}  \\	
&  & $ {d_{z^2},d_{xy},d_{x^2-y^2}}$ & {8.87} & {2.42}  & {2.16} & {0.17} & {1.61}  \\
&  & $ {d}$ & {8.77} & {2.43}  & {2.16} & {0.14} & {1.63}  \\ \addlinespace  $ \rm{WSe_2} $ & $ {1T}$ & $ {d_{z^2}}$ & {8.67} & {0.43}  & {-----} & {-----} & {0.43}  \\	
&  & $ {d_{z^2},d_{xy},d_{x^2-y^2}}$ & {9.09} & {1.10}  & {0.84} & {0.15} & {0.62}  \\
&  & $ {d}$ & {9.09} & {1.43}  & {1.15} & {0.14} & {0.65}  \\ \addlinespace  $ \rm{  } $ & ${1T^{\prime}}$ & ${d_{z^2}}$ & {7.23} & {1.64}  & {-----} & {-----} & {1.64}  \\	
&  & ${d_{z^2},d_{xy},d_{x^2-y^2}}$ & {8.35} & {2.03}  & {1.59} & {0.21} & {1.53}  \\
&  & ${d}$ & {9.17} & {2.41}  & {1.88} & {0.23} & {1.70}  \\ \hline 
$ \rm{  } $ & $ {1H}$ & $ {d_{z^2}}$ & {5.62} & {1.11}  & {-----} & {-----} & {1.11}  \\	
&  & $ {d_{z^2},d_{xy},d_{x^2-y^2}}$ & {6.46} & {1.43}  & {1.26} & {0.09} & {1.00}  \\
&  & $ {d}$ & {6.86} & {1.54}  & {1.33} & {0.09} & {1.08}  \\ \addlinespace  $ \rm{WTe_2} $ & $ {1T}$ & $ {d_{z^2}}$ & {7.21} & {0.39}  & {-----} & {-----} & {0.39}  \\	
&  & $ {d_{z^2},d_{xy},d_{x^2-y^2}}$ & {7.99} & {1.05}  & {0.84} & {0.14} & {0.49}  \\
&  & $ {d}$ & {7.47} & {1.08}  & {0.89} & {0.11} & {0.47}  \\ \addlinespace  $ \rm{  } $ & $ {1T^{\prime}}$ & $ {d_{z^2}}$ & {5.38} & {0.57}  & {-----} & {-----} & {0.57}  \\	
&  & $ {d_{z^2},d_{xy},d_{x^2-y^2}}$ & {6.07} & {0.85}  & {0.59} & {0.12} & {0.68}  \\
&  & $ {d}$ & {7.49} & {1.21}  & {0.85} & {0.17} & {0.89}  \\ 
\end{tabular}
\label{table:UVW-MoW}
\end{ruledtabular}
\end{table*}

\begin{figure*}
\centering
\includegraphics[width=0.85\linewidth]{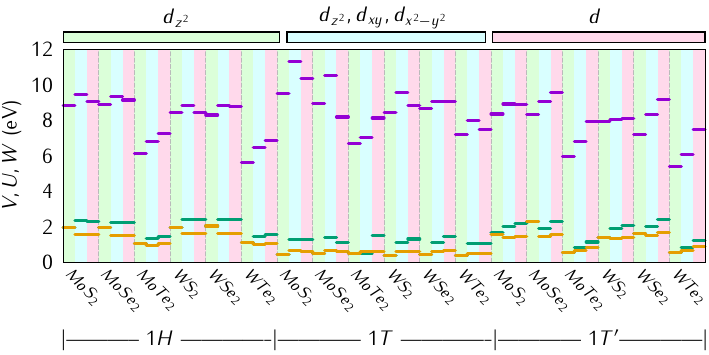}
\caption{(Color online) Comparison of the on-site Coulomb interaction parameters for MX$_2$ 
(M=Mo, W; X=S, Se, Te) in $1H$, $1T$, and $1T'$ structures, considering three different 
correlated subspaces. Purple, green, and orange indicate the values of $V$, $U$, and $W$, 
respectively.}
\label{fig:vuw-MoW}
\end{figure*}

In Table~\ref{table:UVW-MoW}, we present on-site Coulomb interaction parameters, 
including the bare $V$, the partially screened Hubbard-Kanamori parameters $U$, $U'$, and $J$, 
as well as the fully screened $W$. Note that the definitions of $V$ and $W$ are the same as $U$, 
i.e., the average diagonal elements of the Coulomb matrix [see Eq.(\ref{eq.7})]. To facilitate 
a more comprehensive comparison, we visualize the unscreened $V$, partially screened $U$, 
and fully screened $W$ parameters across all compounds and their corresponding phases. This 
graphical representation is presented in Fig.~\ref{fig:vuw-MoW}, enhancing our ability to discern 
the variations and trends in these interactions more effectively.

Let us begin with a discussion of the bare (unscreened) Coulomb interaction. The bare $V$ values 
range from 5.6 to 11.3 eV and depend on the principal quantum number of the $d$ shell, 
chalcogen X, and symmetry of the structure. Our calculated interaction $V$ for M=Mo and 
W with $4d$ and $5d$ orbitals, respectively, are almost 4-5 eV smaller than the bare Coulomb 
interactions of TMDs with $3d$ correlated subspaces \cite{Karbalaee_Aghaee_2022}. Additionally, 
if we focus on a specific subspace, the calculated $V$ parameters for WX$_2$ are smaller than 
MoX$_2$. This is not unexpected, as the bare interaction $V$ generally decreases when moving 
downward in the periodic table from $4d$ TM to $5d$ TM materials, due to the lower degree of contraction of 
the $5d$ wave functions compared to $4d$ and $3d$ ones. Furthermore, the results for the chalcogen 
series MX$_2$ with X=S to Te tend to show a reduction in bare $V$ (with exceptions). One might attribute this to 
the increase in the lattice constant, making the Wannier function more extended or more delocalized. 
This can be seen in Fig.\,\ref{fig:Wannier_orbitals}, which shows the shape of Wannier orbitals for 
Mo(W)X$_2$ compounds. An analysis of the shape of these Wannier functions and the projected band 
structures (see supplementary material) indicates that the coupling of $d$ states to neighboring 
chalcogen $p$ states is not negligible, which leads to delocalization and, therefore, to smaller 
bare interaction $V$ parameters in X=Se, Te compounds compared to X=S. From a symmetry point of view, 
the largest value of bare interaction $V$ is observed in the $1T$ structure. This is consistent 
with the stronger contraction of $d$ wave functions and weaker admixture of chalcogen $p$ with $d$ 
states in the $1T$ structure.

In most of the TMDs considered, the Coulomb interaction is screened efficiently due to the high 
density of M-$d$ and X-$p$ states near the Fermi level $E_F$. As a consequence, the on-site $U$ 
and $W$ values are considerably reduced. The calculated $U$ values for $1H$-MX$_2$, $1T$-MX$_2$, 
and $1T'$-MX$_2$ lie in the range 1.3-2.4 eV, 1.1-1.4 eV, and 0.9-2.4 eV, respectively.
The calculated $U$ and $W$ values depend on the correlated subspace, ground-state electronic 
structure, and chalcogen X atom. In contrast to the bare interaction, the $U$ values are larger 
in the $1H$ structure. This can be attributed to the band gap of the materials in this structure. 
For example, considering a three-orbital correlated subspace, the $U$ values in the $1T$ 
structure are smaller than in $1H$ and $1T'$ due to a metallic screening channel stemming 
from the chalcogen X-$p$ states in the $1T$ and $1T'$ structures (see Fig.\,\ref{fig:interpol-MoW}  
and the projected band structure in the supplementary material).

\begin{table*}[!ht]
\caption{Long-range partially screened Coulomb interaction $U$ for MX$_2$ (M=Mo, W; X=S, Se, Te) 
compounds. $U_{00}$  is the onsite interaction, $U_{01}$ is the nearest neighbor interaction, 
$U_{02}$ is the next nearest neighbor interaction, and so on, up to the sixth 
nearest neighbor interaction.} \label{table:long-MoW}
\centering
\begin{ruledtabular}
\begin{tabular}{cccccccccc}
MX$_2$  & Phase & Orbitals & $U_{00}$(eV)& $U_{01}$(eV) &$U_{02}$(eV) & $U_{03}$(eV)  & $U_{04}$(eV) & $U_{05}$(eV) & $U_{06}$(eV)  \\ \hline					
$ \rm{  } $ & ${1H}$ & ${d_{z^2}}$ & {1.95} & {1.08}  & {0.84} & {0.78} & {0.66} & {0.60} & {0.54}\\	
&  & ${d_{z^2},d_{xy},d_{x^2-y^2}}$ & {2.33} & {1.29}  & {1.01} & {0.93} & {0.79} & {0.72} & {0.65} \\
&  & ${d}$ & {2.30} & {1.29}  & {1.01} & {0.93} & {0.79} & {0.72} & {0.65} \\ \addlinespace  $ \rm{MoS_2} $ & ${1T}$ & ${d_{z^2}}$ & {0.45} & {0.09}  & {0.03} & {0.02} & {0.01} & {0.01} & {0.00}\\	
&  & ${d_{z^2},d_{xy},d_{x^2-y^2}}$ & {1.28} & {0.26}  & {0.09} & {0.07} & {0.03} & {0.02} & {0.01} \\
&  & ${d}$ & {1.27} & {0.27}  & {0.10} & {0.07} & {0.02} & {0.01} & {0.00} \\ \addlinespace  $ \rm{  } $ & ${1T^{\prime}}$ & $ {d_{z^2}}$ & {1.66} & {0.94}  &  {0.89} & {0.71} & {0.71} & {0.67} & {0.59} \\	
&  & ${d_{z^2},d_{xy},d_{x^2-y^2}}$ & {2.00} & {1.13}  & {1.08} & {0.86} & {0.85} & {0.81} & {0.71} \\
&  & ${d}$ & {2.18} & {1.24}  & {1.16} & {0.93} & {0.91} & {0.87} & {0.76}  \\ \hline 
$ \rm{  } $ & ${1H}$ & $ {d_{z^2}}$ & {1.96} & {1.08}  &  {0.86} & {0.80} & {0.68} & {0.63} & {0.56} \\	
&  & ${d_{z^2},d_{xy},d_{x^2-y^2}}$ & {2.26} & {1.25}  & {0.99} & {0.92} & {0.78} & {0.72} & {0.65} \\
&  & ${d}$ & {2.26} & {1.25}  & {0.99} & {0.92} & {0.78} & {0.72} & {0.65}  \\ \addlinespace  $ \rm{MoSe_2} $ & ${1T}$ & $ {d_{z^2}}$ & {0.49} & {0.11}  &  {0.05} & {0.04} & {0.02} & {0.01} & {0.00} \\	
&  & ${d_{z^2},d_{xy},d_{x^2-y^2}}$ & {1.41} & {0.33}  & {0.15} & {0.11} & {0.05} & {0.03} & {0.01} \\
&  & ${d}$ & {1.13} & {0.36}  & {0.17} & {0.12} & {0.05} & {0.03} & {0.01}  \\ \addlinespace  $ \rm{  } $ & ${1T^{\prime}}$ & $ {d_{z^2}}$ & {2.30} & {1.29}  &  {1.25} & {0.99} & {1.01} & {0.94} & {0.83} \\	
&  & ${d_{z^2},d_{xy},d_{x^2-y^2}}$ & {1.89} & {1.06}  & {1.03} & {0.81} & {0.83} & {0.77} & {0.69} \\
&  & ${d}$ & {2.28} & {1.24}  & {1.18} & {0.94} & {0.93} & {0.88} & {0.77}  \\ \hline 
$ \rm{  } $ & ${1H}$ & $ {d_{z^2}}$ & {1.08} & {0.69}  &  {0.55} & {0.52} & {0.44} & {0.41} & {0.37} \\	
&  & ${d_{z^2},d_{xy},d_{x^2-y^2}}$ & {1.35} & {0.86}  & {0.69} & {0.64} & {0.55} & {0.51} & {0.46} \\
&  & ${d}$ & {1.46} & {0.84}  & {0.69} & {0.64} & {0.55} & {0.51} & {0.46}  \\ \addlinespace  $ \rm{MoTe_2} $ & ${1T}$ & $ {d_{z^2}}$ & {0.49} & {0.21}  &  {0.17} & {0.17} & {0.15} & {0.14} & {0.13} \\	
&  & ${d_{z^2},d_{xy},d_{x^2-y^2}}$ & {0.52} & {0.22}  & {0.19} & {0.18} & {0.16} & {0.15} & {0.14} \\
&  & ${d}$ & {1.53} & {0.82}  & {0.66} & {0.62} & {0.53} & {0.49} & {0.44}  \\ \addlinespace  $ \rm{  } $ & ${1T^{\prime}}$ & $ {d_{z^2}}$ & {0.55} & {0.17}  &  {0.16} & {0.10} & {0.09} & {0.09} & {0.07} \\	
&  & ${d_{z^2},d_{xy},d_{x^2-y^2}}$ & {0.85} & {0.27}  & {0.25} & {0.16} & {0.13} & {0.15} & {0.10} \\
&  & ${d}$ & {1.14} & {0.32}  & {0.28} & {0.17} & {0.14} & {0.15} & {0.10}  \\ \hline 
$ \rm{  } $ & ${1H}$ & $ {d_{z^2}}$ & {1.98} & {1.13}  &  {0.88} & {0.81} & {0.68} & {0.62} & {0.56} \\	
&  & ${d_{z^2},d_{xy},d_{x^2-y^2}}$ & {2.42} & {1.38}  & {1.07} & {0.99} & {0.83} & {0.76} & {0.68} \\
&  & ${d}$ & {2.39} & {1.38}  & {1.07} & {0.99} & {0.83} & {0.76} & {0.68}  \\ \addlinespace  $ \rm{WS_2} $ & ${1T}$ & $ {d_{z^2}}$ & {0.40} & {0.09}  &  {0.04} & {0.03} & {0.01} & {0.01} & {0.01} \\	
&  & ${d_{z^2},d_{xy},d_{x^2-y^2}}$ & {1.09} & {0.25}  & {0.10} & {0.08} & {0.04} & {0.02} & {0.01} \\
&  & ${d}$ & {1.31} & {0.40}  & {0.18} & {0.14} & {0.05} & {0.03} & {0.00}  \\ \addlinespace  $ \rm{  } $ & ${1T^{\prime}}$ & $ {d_{z^2}}$ & {1.40} & {0.77}  &  {0.73} & {0.58} & {0.56} & {0.55} & {0.47} \\	
&  & ${d_{z^2},d_{xy},d_{x^2-y^2}}$ & {1.90} & {1.04}  & {0.99} & {0.79} & {0.76} & {0.74} & {0.64} \\
&  & ${d}$ & {2.08} & {1.14}  & {1.06} & {0.86} & {0.81} & {0.80} & {0.68}  \\ \hline 
$ \rm{  } $ & ${1H}$ & $ {d_{z^2}}$ & {2.04} & {1.14}  &  {0.90} & {0.84} & {0.71} & {0.65} & {0.58} \\	
&  & ${d_{z^2},d_{xy},d_{x^2-y^2}}$ & {2.42} & {1.36}  & {1.07} & {0.99} & {0.84} & {0.77} & {0.69} \\
&  & ${d}$ & {2.43} & {1.35}  & {1.07} & {0.99} & {0.84} & {0.77} & {0.69}  \\ \addlinespace  $ \rm{WSe_2} $ & ${1T}$ & $ {d_{z^2}}$ & {0.43} & {0.09}  &  {0.04} & {0.03} & {0.02} & {0.01} & {0.01} \\	
&  & ${d_{z^2},d_{xy},d_{x^2-y^2}}$ & {1.10} & {0.24}  & {0.10} & {0.08} & {0.04} & {0.03} & {0.01} \\
&  & ${d}$ & {1.43} & {0.39}  & {0.18} & {0.13} & {0.05} & {0.03} & {0.00}  \\ \addlinespace  $ \rm{  } $ & ${1T^{\prime}}$ & ${d_{z^2}}$ & {1.64} & {0.95}  & {0.92} & {0.74} & {0.73} & {0.70} & {0.61}\\	
&  & ${d_{z^2},d_{xy},d_{x^2-y^2}}$ & {2.03} & {1.18}  & {1.14} & {0.91} & {0.91} & {0.86} & {0.76} \\
&  & ${d}$ & {2.41} & {1.31}  & {1.24} & {0.99} & {0.97} & {0.93} & {0.81} \\ \hline 
$ \rm{  } $ & ${1H}$ & $ {d_{z^2}}$ & {1.11} & {0.72}  &  {0.58} & {0.54} & {0.46} & {0.42} & {0.38} \\	
&  & ${d_{z^2},d_{xy},d_{x^2-y^2}}$ & {1.43} & {0.93}  & {0.74} & {0.69} & {0.59} & {0.54} & {0.48} \\
&  & ${d}$ & {1.54} & {0.91}  & {0.74} & {0.69} & {0.59} & {0.54} & {0.48}  \\ \addlinespace  $ \rm{WTe_2} $ & ${1T}$ & $ {d_{z^2}}$ & {0.39} & {0.12}  &  {0.06} & {0.05} & {0.03} & {0.02} & {0.01} \\	
&  & ${d_{z^2},d_{xy},d_{x^2-y^2}}$ & {1.05} & {0.32}  & {0.17} & {0.14} & {0.07} & {0.04} & {0.02} \\
&  & ${d}$ & {1.08} & {0.36}  & {0.19} & {0.15} & {0.08} & {0.05} & {0.02}  \\ \addlinespace  $ \rm{  } $ & ${1T^{\prime}}$ & $ {d_{z^2}}$ & {0.57} & {0.21}  &  {0.20} & {0.12} & {0.11} & {0.11} & {0.08} \\	
&  & ${d_{z^2},d_{xy},d_{x^2-y^2}}$ & {0.85} & {0.31}  & {0.29} & {0.18} & {0.16} & {0.17} & {0.12} \\
&  & ${d}$ & {1.21} & {0.37}  & {0.33} & {0.20} & {0.17} & {0.17} & {0.11}  \\ 
\end{tabular}
\end{ruledtabular}
\end{table*}

In the $1T'$ structure, all $d$ states are split due to its lower symmetry compared to the $1T$ structure, 
which significantly modifies the electronic structure \cite{keum2015bandgap}. In this context, the 
full $d$ orbital set is the optimal correlated subspace for capturing the electronic characteristics 
of this structure. The semimetallic behavior of the $1T'$ structure enhances electron-electron 
interactions compared to the $1T$ structure \cite{huang2016controlling,park2020evidence,keum2015bandgap,song2018few}. 
While the $1T'$ structure does not have a band gap, the scarcity of metallic states near 
the Fermi level reduces the contribution of X-$p$ $\longrightarrow $ M-$d$ transitions to screening, resulting 
in a $U$ value that is largely similar to that of the $1H$ structure.

Regardless of the correlated subspace or structural symmetry, electron screening is enhanced in the MTe$_2$ compared to MS$_2$/MSe$_2$, leading to a reduction in both the Coulomb interaction parameters $U$ and $W$. 
For example, in the three-orbital correlated subspace, the $U$ value decreases from 2.33 eV in $1H$-MoS$_2$ 
to 1.35 eV in $1H$-MoTe$_2$. The transition from S to Te in each MX$_2$ system contributes to the determination 
of the $U$ and $W$ parameters through two mechanisms. First, similar to the effect seen in the bare interaction, 
the Wannier localization effect causes a reduction in $U$ and $W$ from S to Te. Second, the shift from S/Se to Te brings 
the X-$p$ states into closer energy proximity with the M-$d$ states, as shown in Fig.\,S1 of the supplementary 
material. This closeness in energy levels translates to a smaller energy difference, which increases the contribution 
of X-$p$ $\longrightarrow $ M-$d$  transitions to the polarization function and as a consequence reduces the $U$ 
and $W$  parameters.  On the other hand, a comparison of MoX$_2$ and WX$_2$ compounds reveals that the screened 
interactions are nearly identical for most of them. This is likely due to the similar atomic radii of Mo and 
W, which leads to a similar degree of localization of the $d$ orbitals in both materials.

\begin{figure} 	
\centering
\captionsetup[subfigure]{justification=justified,singlelinecheck=false,position=top}
\subfloat{\includegraphics[width=1.0\linewidth]{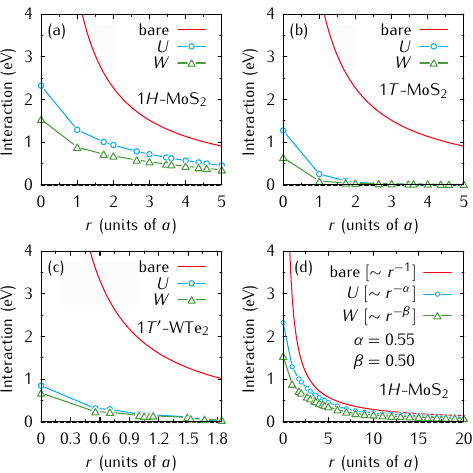}
{\label{long-MoW}}}	
\centering
\caption{(Color online) Distance ($r$) dependence of the partially and fully screened Coulomb 
interaction $U(r)$ and $W(r)$ for  $t_{2g}$ electrons in (a) $1H$-MoS$_2$, (b) 1$T^\prime$-WTe$_2$, 
and (c) $1T$-MoS$_2$. Bare Coulomb interaction $V(r)$ is depicted with a solid line. Panel (d) shows 
the behavior of $U(r)$ and $W(r)$ for $1H$-MoS$_2$ at much larger distances.}
\label{fig:long-MoW}
\end{figure}

So far, we have only considered the on-site Coulomb interaction matrix elements. 
The long-range off-site Coulomb interaction plays an important role in determining 
the phase diagram of 2D materials. The long-range behavior of the screened Coulomb 
interaction is shown in Fig.\,\ref{fig:long-MoW} for four MX$_2$ compounds and 
compared with the unscreened $1/r$ interaction. Table~\ref{table:long-MoW}  reports 
the off-site partially screened Coulomb interaction $U(r)$ for all considered 
Mo- and W-based TMDs as a function of distance up to $r=5a$. As shown in 
Fig.\,\ref{fig:long-MoW}(a), the effective Coulomb interaction in $1H$-MoS$_2$ 
reveals a significant long-range part of $U$. This indicates that the short-range 
interaction is strongly screened, while the long-range interaction is weakly 
screened. Furthermore, due to reduced screening at large distances, in $1H$-MoS$_2$ 
the calculated $U(r)$ approaches the bare interaction $1/r$. The metallic states 
in the $1T$ structure [see Fig.~\ref{fig:long-MoW}(c)] give rise to a significant 
reduction in long-range Coulomb interaction, and it is fully screened at a short 
distance of about 1.5a. In contrast, for the $1H$-MoS$_2$ it takes a considerably 
larger value at a short distance. We find that the $\epsilon(r)=V/U$ ratio for 
$1H$-MoS$_2$ has a strong $r$-dependence, i.e., $\epsilon(r_1)$=3.4, $\epsilon(r_2)$=2.1, 
$\epsilon(r_3)$=1.7, where the distance $r_i$ is given in units of lattice 
constant $a$. This means that the $r$-dependent screening in $1H$-MX$_2$ deviates 
substantially from $1/\epsilon r$, i.e., $U(r)$ and $W(r)$ cannot be expressed by 
a simple static dielectric constant $\epsilon$. The situation is quite similar 
in other considered TMDs with $1H$ structure, where the dielectric constant 
decreases with distance, in agreement with recent experiments. This $r$-dependent 
nonconventional screening explains the large exciton binding energies and deviations 
from the usual hydrogenic Rydberg series of energy levels of the excitonic states 
in semiconducting monolayer TMDs \cite{Chernikov_2014,Qiu_2013}.
Note that the long-range behavior of the $W(r)$ for bulk  MoS$_2$ can be fitted by a static
dielectric constant $\epsilon_{||}=9$  (see Fig.\,S3 in supplementary material), revealing 
a conventional screening in three-dimensional semiconductors. Indeed,  the dielectric constant 
$\epsilon=9$ that we use is very close to the experimental value of 
$\epsilon^{expt.}=10$ \cite{chen2015probing}.

%

%

Most monolayer MoX$_2$ and WX$_2$ materials are stable only in 
the $1H$ structure. However, recent work has shown that 
monolayer of WSe$_2$ can also be grown in 
the $1T'$ structure, which has a topological gap of 129 meV 
as measured by angle-resolved photoemission spectroscopy 
(ARPES) experiments \cite{Chen_2018,Ugeda_2018}.
In the absence of spin-orbit coupling, $1T'$-WX$_2$ is a 
semimetal, with two bands meeting at the Fermi level (E$_F$), 
similar to graphene. However, the valence band maximum in 
$1T'$-WX$_2$ has a flattened non-parabolic shape along the 
$\Gamma$-Y direction, while the conduction and valence bands 
in graphene meet at the K point with linear dispersion. 
From the point of view of screening, electrons in linear bands behave like electrons 
in an insulator, and the long-range part of the effective 
Coulomb interaction is not screened well \cite{Hwang_2007,Wehling_2011,Kotov_2012}. 
In the $1T'$ structure of MX$_2$ compounds, the screening also 
turns out to be non-local, even though the bands do not 
have purely linear dispersion. For example, in the full $d$ subspace 
of MoS$_2$, the non-local sixth neighbor interaction $U_{06}$ 
for the $1T'$ structure is 0.76 eV, which is even larger than the 
corresponding value for the semiconducting $1H$ structure ($U_{06}$ = 0.65 eV).
These findings have important implications for our understanding 
of the interaction effects in 2D materials. It suggests that the 
relationship between long-range screening and linear dispersion 
is not straightforward and that other factors such as the density of correlated states around E$_F$ and the overlap of the $p$ orbitals with these states also play a role.

\subsection{Coulomb interaction parameters: MX$_2$ (M=Nb, Ta; X=S, Se, Te)}
\label{}

\begin{figure*} 
\centering		
\includegraphics[width=0.85\linewidth]{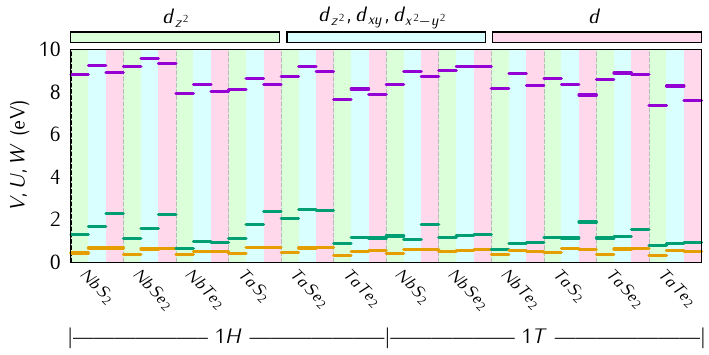}
\caption{(Color online) Comparison of the on-site Coulomb interaction parameters for MX$_2$ (M=Nb, Ta; X=S, Se, Te) in $1H$ and $1T$ structures, considering three different correlated subspaces. Purple, green, and orange indicate the values of $V$, $U$, and $W$, respectively.}
\label{fig:vuw-NbTa}
\end{figure*}

The 2D Mo(W)X$_2$ compounds discussed in the preceding section can have different ground states, 
ranging from semiconducting to semimetallic to metallic, depending on the crystal structure. In 
this section, we investigate the screening of the Coulomb interaction in Nb- and Ta-based MX$_2$ 
compounds (where X=S, Se, Te). Nb and Ta atoms have one fewer electron than Mo and W atoms. Their 
electronic structures show different metallic behaviors. In the $1H$ structure, these materials
show cold metallic behavior, while in the $1T$ structure they are normal metals (see Fig.\,\ref{fig0} and Fig.\,\ref{fig:interpol-NbTa} ). As can be seen from the Wannier interpolated band structures in Fig.\,\ref{fig:interpol-NbTa} 
a minimal one-orbital low-energy model may be sufficient for investigating transport properties 
in the $1H$ structure of these materials. However, the $1T$ structure requires a more 
comprehensive three-orbital low-energy model. For completeness, we provide Coulomb interaction 
parameters for three correlated subspaces: a single orbital ($d_{z^2}$), a three-orbitals ($d_{z^2}$ + 
$d_{xy}$ + $d_{x^2-y^2}$), and the full $d$ shell of the TM atom. The calculated Coulomb 
interaction parameters, including the bare $V$, Hubbard-Kanamori parameters ($U$, $U'$, and $J$), 
and the fully screened $W$, are systematically presented in Table~\ref{table:VUW-NbTa} and Fig.\,\ref{fig:vuw-NbTa} provides a visual 
representation of the trends of $V$, $U$, and $W$ parameters across all compounds and their 
respective phases, to facilitate the comparison.

\begin{table*}[!ht]
\caption{Bare Coulomb interaction $V$, partially screened Hubbard-Kanamori parameters [$U$,  $U^{\prime}$, $J$ (in eV)] and fully screened  $W$ (in eV) for $d$ orbitals of TMs in  MX$_2$ (M=Nb, Ta; X=S, Se, Te) compounds.}
\centering
\begin{ruledtabular}
\begin{tabular}{cccclccc}
MX$_2$  &Phase &  Orbitals & $V^{}$(eV)&$U^{}$(eV)& ${U^{\prime}}(eV)^{}$  & $J(eV)^{}$ & $W^{}$(eV)  \\ \hline	
$ \rm{  } $ & ${1H}$ & ${d_{z^2}}$ & {8.84} & {1.29}  & {-----} & {-----} & {0.43}  \\	
&  & ${d_{z^2},d_{xy},d_{x^2-y^2}}$ & {9.25} & {1.69}  & {1.39} & {0.15} & {0.67}  \\
&  & ${d}$ & {8.92} & {2.30}  & {2.01} & {0.13} & {0.67}  \\ \addlinespace  $ \rm{NbS_2} $ & ${1T}$ & ${d_{z^2}}$ & {8.36} & {1.23}  & {-----} & {-----} & {0.41}  \\	
&  & ${d_{z^2},d_{xy},d_{x^2-y^2}}$ & {8.99} & {1.08}  & {0.83} & {0.14} & {0.61}  \\
&  & ${d}$ & {8.74} & {1.77}  & {1.52} & {0.13} & {0.61}  \\ \addlinespace  \hline \addlinespace
				
$ \rm{  } $ & ${1H}$ & ${d_{z^2}}$ & {9.21} & {1.11}  & {-----} & {-----} & {0.37} \\	
&  & ${d_{z^2},d_{xy},d_{x^2-y^2}}$ & {9.58} & {1.59}  & {1.29} & {0.17} & {0.62}  \\
&  & ${d}$ & {9.35} & {2.25}  & {1.97} & {0.14} & {0.64}  \\ \addlinespace  $ \rm{NbSe_2} $ & $ {1T}$ & ${d_{z^2}}$  & {9.01} & {1.16}  & {-----} & {-----} & {0.52} \\	
&  & ${d_{z^2},d_{xy},d_{x^2-y^2}}$ & {9.19} & {1.25}  & {0.96} & {0.16} & {0.56}  \\
&  & ${d}$ & {9.19} & {1.29}  & {1.03} & {0.14} & {0.58}  \\ \addlinespace  \hline \addlinespace
				
$ \rm{  } $ & $ {1H}$ & ${d_{z^2}}$  & {7.93} & {0.65}  & {-----} & {-----} & {0.36} \\	
&  & ${d_{z^2},d_{xy},d_{x^2-y^2}}$ & {8.36} & {0.96}  & {0.71} & {0.12} & {0.49}  \\
&  & ${d}$ & {8.05} & {0.94}  & {0.70} & {0.10} & {0.49}  \\ \addlinespace  $ \rm{NbTe_2} $ & $ {1T}$ & ${d_{z^2}}$  & {8.18} & {0.60}  & {-----} & {-----} & {0.35} \\	
&  & ${d_{z^2},d_{xy},d_{x^2-y^2}}$ & {8.86} & {0.86}  & {0.59} & {0.13} & {0.53}  \\
&  & ${d}$ & {8.30} & {0.93}  & {0.70} & {0.11} & {0.50}  \\ \addlinespace  \hline \addlinespace
				
$ \rm{  } $ & ${1H}$ & ${d_{z^2}}$ & {8.12} & {1.13}  & {-----} & {-----} & {0.41} \\	
&  & ${d_{z^2},d_{xy},d_{x^2-y^2}}$ & {8.66} & {1.76}  & {1.47} & {0.16} & {0.69}  \\
&  & ${d}$ & {8.34} & {2.40}  & {2.12} & {0.12} & {0.70}  \\ \addlinespace  $ \rm{TaS_2} $ & $ {1T}$ & ${d_{z^2}}$  & {8.63} & {1.15}  & {-----} & {-----} & {0.46} \\	
&  & ${d_{z^2},d_{xy},d_{x^2-y^2}}$ & {8.35} & {1.14}   & {0.90} & {0.14} & {0.65}  \\
&  & ${d}$ & {7.87} & {1.89}  & {1.66} & {0.12} & {0.61}  \\ \addlinespace  \hline \addlinespace
				
$ \rm{  } $ & $ {1H}$ & ${d_{z^2}}$  & {8.74} & {2.05}  & {-----} & {-----} & {0.45} \\	
&  & ${d_{z^2},d_{xy},d_{x^2-y^2}}$ & {9.19} & {2.47}  & {2.18} & {0.19} & {0.67}  \\
&  & ${d}$ & {8.96} & {2.45}  & {2.17} & {0.14} & {0.69}  \\ \addlinespace  $ \rm{TaSe_2} $ & $ {1T}$ & ${d_{z^2}}$  & {8.58} & {1.14}  & {-----} & {-----} & {0.38} \\	
&  & ${d_{z^2},d_{xy},d_{x^2-y^2}}$ & {8.90} & {1.21}  & {0.95} & {0.16} & {0.62}  \\
&  & ${d}$ & {8.85} & {1.52}  & {1.26} & {0.14} & {0.64}  \\ \addlinespace  \hline \addlinespace
				
$ \rm{  } $ & $ {1H}$ & ${d_{z^2}}$  & {7.67} & {0.90}  & {-----} & {-----} & {0.33} \\	
&  & ${d_{z^2},d_{xy},d_{x^2-y^2}}$ & {8.15} & {1.17}  & {0.92} & {0.15} & {0.52}  \\
&  & ${d}$ & {7.88} & {1.14}  & {0.88} & {0.11} & {0.53}  \\ \addlinespace  $ \rm{TaTe_2} $ & $ {1T}$ & ${d_{z^2}}$  & {7.36} & {0.78}  & {-----} & {-----} & {0.30} \\	
&  & ${d_{z^2},d_{xy},d_{x^2-y^2}}$ & {8.29} & {0.87}   & {0.61} & {0.12} & {0.54}  \\
&  & ${d}$ & {7.62} & {0.94}  & {0.72} & {0.11} & {0.50}  \\ \addlinespace

\end{tabular}
\label{table:VUW-NbTa}
\end{ruledtabular}
\end{table*}

We find that the calculated $V$ parameters are slightly smaller than the same parameters in Mo(W)X$_2$
compounds. This is because the Nb (Ta) atom has one electron less than the Mo (W) atom (less nuclear charge), and thus 
the wave functions are less contracted, leading to less localized Wannier functions and as a consequence 
slightly smaller $V$ parameters. This is consistent with the findings of previous studies \cite{TM_cRPA,Karbalaee_Aghaee_2022,Neroni_2019}.
We also compare the $V$ parameters in Nb(Ta)X$_2$ compounds with the corresponding ones in elementary 
transition metals. We find that the $V$ values for the considered 4$d$/5$d$ TMDs compounds, namely NbX$_2$ 
and TaX$_2$, are almost 2 eV smaller than the corresponding ones in elementary Nb or Ta bulk systems \cite{TM_cRPA}. 
This is likely due to the coupling to neighboring chalcogen $p$ states in TMDs. The coupling to $p$ states 
makes the TM $d$ orbitals spill into the $p$ states, giving rise to delocalization and, therefore, to 
smaller $V$ parameters. This effect is also reflected in the shape of the Wannier orbitals shown in Fig.\,\ref{fig:Wannier_orbitals} and in the projected band structures (see  supplementary material).  An analysis of the shape of the Wannier orbitals
reveals that for any system in which the overlap of the $d$ and $p$ orbitals is stronger, the Wannier functions spread to the nearest neighboring atoms.
Moreover as shown in Fig.\,S4, the coupling to neighboring chalcogen $p$ states gets
stronger in Te-based TMDs, which makes the Wannier functions more extended.

Similarly to the bare Coulomb interaction parameters discussed above, the Wannier localization effect 
is also important for the partially screened $U$ and fully screened $W$ parameters. The $U$ ($W$) parameter 
tends to decrease from Mo/W to Nb/Ta TMDs systems because the Wannier functions are less localized 
in Nb/Ta TMDs due to smaller value of nuclear charge.  In addition, for the $1H$ 
structure of MX$_2$ (M=Nb, Ta), a flat half-filled band with mainly $d_{z^2}$ character increases 
screening substantially. This results in larger $V-W$ and $V-U$ differences compared to the corresponding 
differences for the $1H$-MX$_2$ (M=Mo, W) compounds. Note that although the interaction parameter $U$ is 
smaller in metallic MX$_2$ (M=Nb, Ta) compounds than in $4d/5d$ elementary TMs and other materials 
containing $4d/5d$ transition metals, such as MXenes ($U=3.5-4$ eV) \cite{Hadipour_2019}, the narrow 
bands with $d_{z^2}$, $t_{2g}$, and $e_g$ character in TMDs result in a larger $U/W_b$ correlation strength.

Strong interactions between localized electrons can lead to a transition from 
a metal to an insulator and trigger a magnetic phase change \cite{Xu_2014,G_ller_2016,Lin_2019,Hadipour_2019-s}. 
Magnetic ordering and Mott phases have been experimentally observed in layered materials containing 
transition metal (TM) atoms, such as the TM halides CrI$_3$  \cite{Huang_2017,Seyler_2017}, VI$_3$ 
\cite{Tian_2019,yekta_2021}, the 3d transition metal dichalcogenides (TMDs) $2H$-VX$_2$ 
\cite{Wang_2021,Karbalaee_Aghaee_2022}, and the TM phosphorous trichalcogenides NiPX$_3$ (where 
X=S and Se) \cite{Kim_2018}. All of these materials have flat bands in their energy spectra. In the 
case of $2H$-NbX$_2$ and $2H$-TaX$_2$, there is an isolated, half-filled low-energy band with 
predominantly $d_{z^2}$ character resulting in a large $U/W_b$ ratio. 
Although the $1H$ structure of NbX$_2$ and TaX$_2$ 
has stronger correlations due to a single, narrow $d_{z^2}$  band, Mott-like insulating behavior 
has been observed in the reconstructed $1T$ lattice.

Experimentally, compounds such as TaS$_2$\cite{Butler_2020}, NbSe$_2$ \cite{wang2017high}, and 
TaSe$_2$ \cite{nakata2021robust}  exhibit metallic behavior at high temperatures in the 1$T$ crystal 
structure. However, as the temperature is lowered, they undergo a transition to an insulating phase 
characterized by a distinct atomic rearrangement pattern, known as the "star-of-David" motif 
\cite{wang2020band,petkov2022atomic,liu2021charge,kratochvilova2017low}. For example, the 
reconstructed 1$T$-TaS$_2$ monolayer has a Mott gap of 0.45 eV \cite{Lin_2019}. This lattice 
reconstruction coincides with a commensurate CDW (CCDW) transition.

\begin{table}[!ht]
\caption{The on-site effective Coulomb interaction $U^{\mathrm{eff}}_{00}$  and the 
off-site interaction $U^{\mathrm{eff}}_{01}$  are calculated for a specific atomic 
rearrangement pattern within the 1$T$ crystal structure, known as the star-of-David 
configuration.} \label{table:U_eff}
\centering
\begin{ruledtabular}
\begin{tabular}{lccll}
MX$_2$  &Phase & Orbitals & $U^\mathrm{eff}_{00}$(eV)& $U^{\mathrm{eff}}_{01}$(eV)  \\ \hline	
				
NbS$_2$  & $1T$ &  $d$ & 0.71 & 0.40  \\  \addlinespace
NbSe$_2$ & $1T$ &  $d$ & 0.27 & 0.06 \\  \addlinespace
NbTe$_2$ & $1T$ &  $d$ & 0.18 & 0.04 \\  \addlinespace
TaS$_2$ & $1T$  &  $d$ & 0.83 & 0.47 \\\addlinespace
TaSe$_2$ & $1T$ &  $d$ & 0.35 & 0.07 \\ \addlinespace
TaTe$_2$ & $1T$ &  $d$ & 0.18 & 0.04 \\
\end{tabular}
\label{table-U_eff}
\end{ruledtabular}
\end{table}

The effective Coulomb interaction $U^{\mathrm{eff}}$ for the star-of-David configuration in  MX$_2$ lattices 
has been reported in the literature by some authors \cite{Darancet_2014,Kamil_2018,Kim_2023}. Following the same procedure  
[see Eq.(\ref{eq.10})] and using the results of the undistorted 1$T$ structure, we estimated the on-site and off-site Coulomb 
interactions. The results are presented in Table~\ref{table-U_eff}. In the cases of NbS$_2$ and TaS$_2$, 
we obtain $U^{\mathrm{eff}}_{00}$ values of 0.71 eV and 0.83 eV, respectively. This is 0.18 eV larger than the value 
of 0.65 eV reported for TaS$_2$ in a previous study using the same approach \cite{Kim_2023}. The on-site interaction 
$U^{\mathrm{eff}}_{00}$ is large enough compared to the bandwidth $W_b$=0.02 eV \cite{Darancet_2014,Kim_2023} which 
induces a robust Mott insulator in CCDW 1$T$-TaS$_2$. Moreover, $U^{\mathrm{eff}}_{00}$ for TaSe$_2$ is also sizable, revealing 
the importance of the correlation in other MX$_2$ which have not been quantitatively studied before.

The stabilization of the distorted CCDW 1$T$-MX$_2$ (M=Nb, Ta) not only increases 
the ratio of the on-site effective Coulomb interaction $U^{\mathrm{eff}}_{00}$ to 
the bandwidth $W_b$, but also leads to an increase in the ratio of the long-range 
interaction $U^{\mathrm{eff}}_{01}$ to $U^{\mathrm{eff}}_{00}$ up to 0.56. This could potentially lead 
to the emergence of exotic quantum phenomena. The obtained on-site $U^{\mathrm{eff}}_{00}$ 
and off-site $U^{\mathrm{eff}}_{01}$  Coulomb interaction parameters are important 
for use in extended model Hamiltonians and methods beyond DFT, such as DFT+$U$ and 
DFT plus dynamical mean-field theory (DFT+DMFT), which is used to obtain a reliable 
electronic structure, Mott gap, and other properties.

\begin{figure} 	
\centering
\captionsetup[subfigure]{justification=justified,singlelinecheck=false,position=top}
\subfloat{\includegraphics[width=1.0\linewidth]{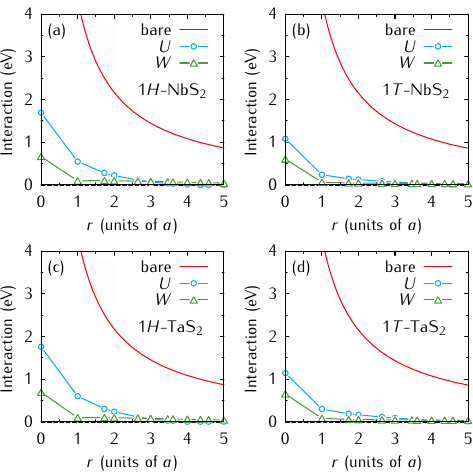}
{\label{long-NbTa}}}
\centering
\caption{(Color online) Distance ($r$) dependence of the partially screened Coulomb interaction 
$U(r)$ and fully screened Coulomb interaction $W(r)$ between $t_{2g}$ electrons in (a) 1$H$-NbS$_2$, 
(b) 1$T$-NbS$_2$ (c) 1$H$-TaS$_2$, and (d) 1$T$-TaS$_2$. The bare Coulomb interaction $V(r)$ is depicted 
with a solid line.}
\label{fig:long-NbTa}
\end{figure}

We have discussed the on-site Coulomb interaction parameters in Nb(Ta)X$_2$ 
(where X=S, Se, Te) compounds.  As in Mo(W)X$_2$ compounds, the screening of the 
long-range Coulomb interaction is nonconventional in these compounds. Figure~\ref{fig:long-NbTa} 
shows the calculated partially and fully screened Coulomb interaction parameters 
$U$ and $W$ for the $1H$ and $1T$ phases of NbS$_2$ and TaS$_2$ compounds as a 
function of distance up to $r=5a$. Despite the metallic nature of these compounds, 
the long-range Coulomb interactions do not screen well, and as a consequence, 
long-range interactions remain sizable, as shown in Table~\ref{table-long-NbTa}. 
The important consequence of this reduced screening and sizable long-range Coulomb 
interaction in metallic systems is the appearance of CDW order and the existence 
of intriguing plasmons in TMDs. The latter will be discussed in detail below. Note 
that CDW appears to be stronger in the single-layer form of 1$T$-MX$_2$ 
\cite{Xi_2015,Ritschel_2015,chen2020strong,Lee_2021}. The sizable $U_{0n}$ in the 
1$T$ structure indicates that the electron-electron interaction is one of the main 
reasons for the occurrence of CDW order in TMDs. Although the value of off-site 
$U_{0n}$ in the 1$T$ structure is not negligible, it is not as large as the 
corresponding one in the 1$H$ structure. It would be interesting 
to take off-site $U_{0n}$ into account within the extended Hubbard model and 
investigate whether CDW order occurs for TaX$_2$ and NbX$_2$ or not.

\begin{table*}[!ht]
\caption{Long-range partially screened Coulomb interaction $U$ for MX$_2$ (M=Nb, Ta; X=S, Se, Te) 
compounds. $U_{00}$  is the onsite interaction, $U_{01}$ is the nearest neighbor interaction, 
$U_{02}$ is the next nearest neighbor interaction, and so on, up to the sixth nearest neighbor interaction.} \label{table:long-NbTa}
\centering
\begin{ruledtabular}
\begin{tabular}{cccccccccc}
MX$_2$  &Phase & Orbitals & $U_{00}$(eV)& $U_{01}$(eV) &$U_{02}(eV)$ & $U_{03}$(eV)  & $U_{04}(eV)$  & $U_{05}(eV)$  & $U_{06}(eV)$  \\ \hline	
				
			$ \rm{  } $ & ${1H}$ & ${d_{z^2}}$ & {1.29} & {0.42}  & {0.22} & {0.17} & {0.09} & {0.06} & {0.03} \\	
			&  & ${d_{z^2},d_{xy},d_{x^2-y^2}}$ & {1.69} & {0.55}  & {0.28} & {0.22} & {0.11} & {0.07} & {0.03} \\
			&  & ${d}$ & {2.30} & {1.23}  & {0.97} & {0.90} & {0.76} & {0.69} & {0.62} \\ \addlinespace  $ \rm{NbS_2} $ & ${1T}$ & ${d_{z^2}}$ & {1.23} & {0.27}  & {0.16} & {0.14} & {0.10} & {0.08} & {0.06} \\	
			&  & ${d_{z^2},d_{xy},d_{x^2-y^2}}$ & {1.08} & {0.24}  & {0.14} & {0.13} & {0.09} & {0.07} & {0.05} \\
			&  & ${d}$ & {1.77} & {0.79}  & {0.61} & {0.57} & {0.48} & {0.45} & {0.40} \\ \addlinespace  \hline \addlinespace
			
			$ \rm{  } $ & ${1H}$ & ${d_{z^2}}$ & {1.11} & {0.34}  & {0.17} & {0.14} & {0.07} & {0.05} & {0.02} \\	
			&  & ${d_{z^2},d_{xy},d_{x^2-y^2}}$ & {1.59} & {0.48}  & {0.25} & {0.20} & {0.10} & {0.06} & {0.03} \\
			&  & ${d}$ & {2.25} & {1.19}  & {0.95} & {0.89} & {0.76} & {0.70} & {0.63}  \\ \addlinespace  $ \rm{NbSe_2} $ & ${1T}$ & ${d_{z^2}}$ & {1.16} & {0.28}  &  {0.18} & {0.15} & {0.11} & {0.09} & {0.06} \\	
			&  & ${d_{z^2},d_{xy},d_{x^2-y^2}}$ & {1.12} & {0.27}  & {0.17} & {0.15} & {0.10} & {0.08} & {0.06}\\
			&  & ${d}$ & {1.29} & {0.32}  & {0.18} & {0.15} & {0.09} & {0.07} & {0.05} \\ \addlinespace  \hline \addlinespace
			
			$ \rm{  } $ & ${1H}$ & ${d_{z^2}}$ & {0.65} & {0.16}  &  {0.09} & {0.08} & {0.05} & {0.03} & {0.02} \\	
			&  & ${d_{z^2},d_{xy},d_{x^2-y^2}}$ & {0.96} & {0.24}  & {0.14} & {0.11} & {0.07} & {0.05} & {0.03}\\
			&  & ${d}$ & {0.94} & {0.25}  & {0.14} & {0.11} & {0.07} & {0.05} & {0.03} \\ \addlinespace  $ \rm{NbTe_2} $ & ${1T}$ & ${d_{z^2}}$ & {0.60} & {0.09}  &  {0.05} & {0.04} & {0.03} & {0.02} & {0.02} \\	
			&  & ${d_{z^2},d_{xy},d_{x^2-y^2}}$ & {0.86} & {0.13}  & {0.07} & {0.06} & {0.04} & {0.03} & {0.02}\\
			&  & ${d}$ & {0.93} & {0.20}  & {0.11} & {0.09} & {0.05} & {0.04} & {0.03} \\ \addlinespace  \hline \addlinespace
			
			$ \rm{  } $ & ${1H}$ & ${d_{z^2}}$ & {1.13} & {0.39}  & {0.20} & {0.15} & {0.07} & {0.05} & {0.02}\\	
			&  & ${d_{z^2},d_{xy},d_{x^2-y^2}}$ & {1.76} & {0.60}  & {0.31} & {0.24} & {0.12} & {0.07} & {0.03} \\
			&  & ${d}$ & {2.40} & {1.32}  & {1.03} & {0.96} & {0.80} & {0.73} & {0.65} \\ \addlinespace  $ \rm{TaS_2} $ & ${1T}$ & ${d_{z^2}}$ & {1.15} & {0.30}  &  {0.19} & {0.17} & {0.12} & {0.09} & {0.07} \\	
			&  & ${d_{z^2},d_{xy},d_{x^2-y^2}}$ & {1.14} & {0.30}  & {0.19} & {0.16} & {0.11} & {0.09} & {0.06}\\
			&  & ${d}$ & {1.89} & {0.95}  & {0.73} & {0.68} & {0.58} & {0.53} & {0.48} \\ \addlinespace  \hline \addlinespace
			
			$ \rm{  } $ & ${1H}$ & ${d_{z^2}}$ & {2.05} & {1.09}  &  {0.86} & {0.80} & {0.68} & {0.62} & {0.56} \\	
			&  & ${d_{z^2},d_{xy},d_{x^2-y^2}}$ & {2.47} & {1.31}  & {1.04} & {0.96} & {0.82} & {0.75} & {0.67}\\
			&  & ${d}$ & {2.45} & {1.31}  & {1.03} & {0.96} & {0.82} & {0.75} & {0.67} \\ \addlinespace  $ \rm{TaSe_2} $ & ${1T}$ & ${d_{z^2}}$ & {1.14} & {0.28}  &  {0.17} & {0.15} & {0.10} & {0.08} & {0.06} \\	
			&  & ${d_{z^2},d_{xy},d_{x^2-y^2}}$ & {1.21} & {0.29}  & {0.18} & {0.15} & {0.11} & {0.09} & {0.06}\\
			&  & ${d}$ & {1.52} & {0.44}  & {0.24} & {0.20} & {0.12} & {0.09} & {0.06} \\ \addlinespace  \hline \addlinespace
			
			$ \rm{  } $ & ${1H}$ & ${d_{z^2}}$ & {0.90} & {0.26}  &  {0.14} & {0.11} & {0.06} & {0.04} & {0.02} \\	
			&  & ${d_{z^2},d_{xy},d_{x^2-y^2}}$ & {1.17} & {0.34}  & {0.18} & {0.14} & {0.08} & {0.05} & {0.03}\\
			&  & ${d}$ & {1.14} & {0.33}  & {0.17} & {0.14} & {0.07} & {0.05} & {0.03} \\ \addlinespace  $ \rm{TaTe_2} $ & ${1T}$ & ${d_{z^2}}$ & {0.78} & {0.12}  &  {0.06} & {0.05} & {0.03} & {0.03} & {0.02} \\	
			&  & ${d_{z^2},d_{xy},d_{x^2-y^2}}$ & {0.87} & {0.13}  & {0.07} & {0.06} & {0.04} & {0.03} & {0.02}\\
			&  & ${d}$ & {0.94} & {0.21}  & {0.11} & {0.09} & {0.05} & {0.04} & {0.03} \\ \addlinespace

			\end{tabular}
			\label{table-long-NbTa}
		\end{ruledtabular}
	\end{table*}

\begin{figure*}[!ht]	
\captionsetup[subfigure]{justification=justified,singlelinecheck=false,position=top}		
\subfloat{
\includegraphics[width=1.0\linewidth]{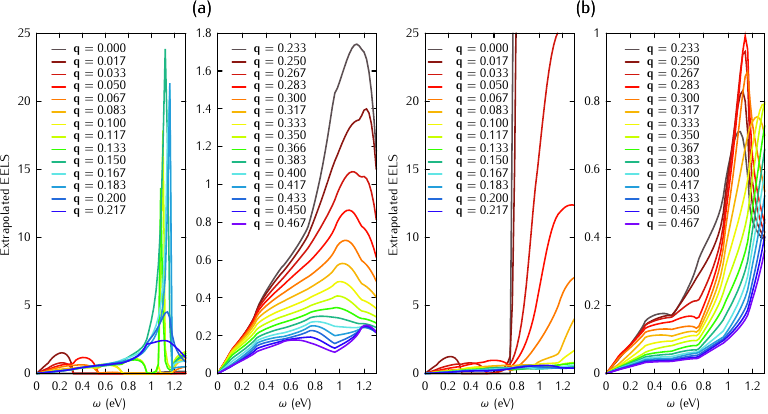}
{\label{LF}}}			
\centering
\vspace{-0.2 cm}
\caption{Extrapolated electron energy loss spectra for selected wave vectors along the 
$\Gamma$-M direction for (a) 1$H$-TaS$_2$, (b) 1$T$-TaS$_2$.}
\label{fig:LF}
\end{figure*}

\begin{figure} 	
\captionsetup[subfigure]{justification=justified,singlelinecheck=false,position=top}
\subfloat{	\includegraphics[width=1.0\linewidth]{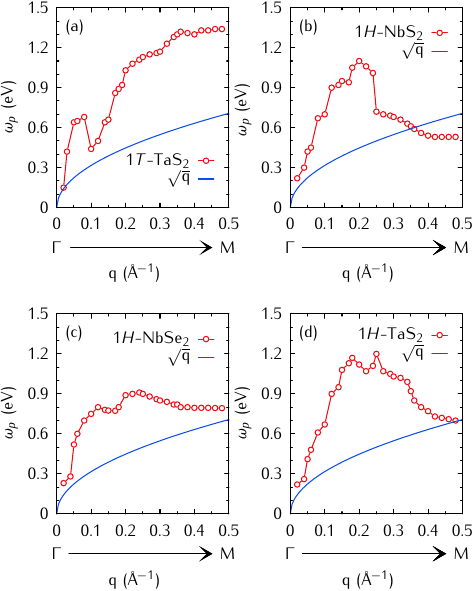}
{\label{fig1:plasm-TaS2-1T}}}
\centering
\caption{Plasmon dispersion along the high symmetry line $\Gamma$-M in the 2D Brillouin zone
for (a) 1$T$-TaS$_2$, (b) 1$H$-NbS$_2$, (c) 1$H$-NbSe$_2$, and (d)  1$H$-TaS$_2$.}
\label{fig:plasm-freq}
\end{figure}

The plasmon dispersion of monolayer TMDs has been studied extensively in recent 
years \cite{Song_2021,Cudazzo_2016,da2020universal,da_Jornada_2020}. While the 
dispersion of plasmons in three-dimensional bulk systems commences with finite 
energy at a wavevector of $q=0$, conventional plasmons in 2D metallic systems follow 
a $\omega_p \propto \sqrt{q}$ relationship. To ascertain plasmon excitations, we 
identify the peaks in the loss function $L(\mathbf{q},\omega) = -\text{Im}(1/\epsilon_m(\mathbf{q},\omega))$, 
wherein $\epsilon_m(\mathbf{q},\omega) = 1/\epsilon_{00}^{-1}(\mathbf{q},\omega)$ 
represents the macroscopic dielectric function. Fig.~\ref{fig:plasm-freq} displays 
the calculated plasmon dispersion for the monolayers of four compounds: 1$T$-TaS$_2$, 
1$H$-NbS$_2$, 1$H$-NbSe$_2$, and 1$H$-TaS$_2$ and compared with the  $\sqrt{q}$ 
plasmon dispersion. Surprisingly, materials of the form MX$_2$ (M=Nb, Ta) deviate 
from the anticipated $\sqrt{q}$ plasmon dispersion observed in conventional 2D metallic 
systems. Particularly striking is the nearly linear dispersion of plasmons in the 
monolayer metallic 1$H$-NbS$_2$ at small $\mathbf{q}$, which transitions to a nearly 
dispersionless behavior within an intermediate range of wavevectors ($\mathbf{q}=0.1$ 
\AA$^{-1}$ to 0.3 \,\AA$^{-1}$). Moreover, we observe a negative slope in the 
dispersion relation for 1$H$-MX$_2$ at larger $\mathbf{q}$. This deviation from the 
$\sqrt{q}$ behavior is most pronounced in the 1$H$ structure due to its comparatively 
greater long-range interaction compared to the 1$T$ structure. Notably, these findings 
align well with electron energy loss spectroscopy (EELS) measurements conducted 
by other researchers, who noted a negative slope in the plasmon dispersion of 1$H$-TaS$_2$, 
1$H$-TaSe$_2$, and 1$H$-NbSe$_2$ \cite{van_Wezel_2011}.  The negative plasmon dispersion 
in TMDs has been attributed to the strong electron-electron interactions in 
these materials. These interactions can lead to the formation of collective charge 
fluctuations, which can couple to the plasmons and modify their dispersion. The 
results of our study provide new insights into the plasmon properties of TMDs. 
These insights could be used to design TMD-based plasmonic devices with novel 
functionalities.

\subsection{Frequency dependency of screened Coulomb interaction}
\label{sec:frequency dependence}

\begin{figure} 
\captionsetup[subfigure]{justification=justified,singlelinecheck=false,position=top}		
\subfloat{
\includegraphics[width=1.0\linewidth]{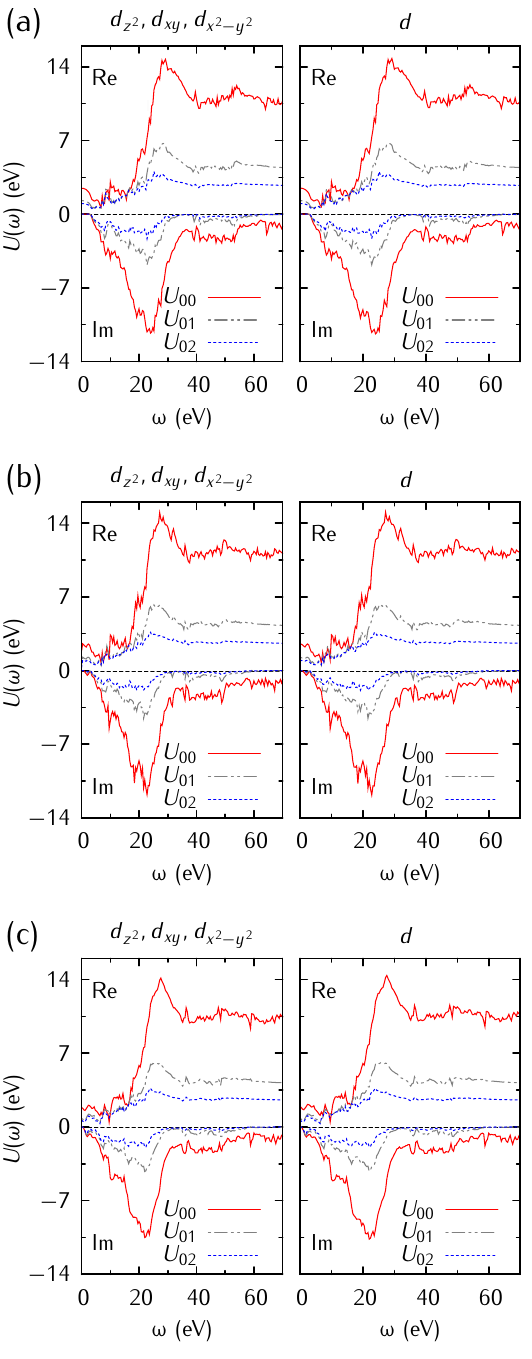}
{\label{freq-MoS2-1H}}}
\centering
\caption{(Color online) Frequency dependence of the on-site and off-site Coulomb 
interaction parameters $U(\omega)$ for (a) 1$H$-MoS$_2$, (b) 1$H$-NbS$_2$,  and (c) 1$T$-NbS$_2$. The real and imaginary parts of $U(\omega)$
for different correlated subspaces are presented individually.}
\label{fig:freq}
\end{figure}

In this section, we investigate the frequency-dependent behavior of the 
partially $U(\omega)$. We focus on the semiconducting 1$H$-MoS$_2$, metallic 1$H$-NbS$_2$, 
and metallic 1$T$-NbS$_2$ compounds. We analyze these materials with respect 
to two distinct correlated subspaces. The real and imaginary components of the 
computed on-site interaction  $U_{00}(\omega)$, as well as the interactions
of the first and second nearest neighbors, namely  $U_{01}(\omega)$ and  
$U_{02}(\omega)$, are shown in Fig.~\ref{fig:freq} for these materials. 
Given the structural and compositional similarities among them, the 
frequency-dependent  behaviors of the screened Coulomb interactions 
exhibit comparable trends across all three systems.

In the case of 1$H$-MoS$_2$, the $U(\omega)$ profile displays smooth behavior with minor fluctuations 
up to a frequency of 15 eV. Beyond this point, it experiences linear growth, ultimately peaking at 
the plasmon frequency of approximately 25 eV (which is also evident as a peak in the imaginary 
part of $U(\omega)$). Subsequently, as the frequency increases further, it asymptotically approaches 
the static value of 9.5 eV characteristic of 1$H$-MoS$_2$. 

For 1$H$-NbS$_2$ and 1$T$-NbS$_2$, both metallic systems, the Coulomb interactions exhibit more pronounced 
variations at lower frequencies. These strong variations can be attributed to the narrow bands around the
Fermi level in these materials.  Notably, the frequency dependency of $U_{00}(\omega)$ shows similar 
trends to the $U_{01}(\omega)$ and $U_{02}(\omega)$ cases. The decline in $U(\omega)$ 
at lower frequencies, around 5 eV, in metallic systems arises from effective screening influenced by $d$ 
states around the Fermi level, which forms a bandwidth of approximately 5 eV. This behavior extends to 
the off-site Coulomb interactions (nearest-neighbor $U_{01}(\omega)$  and next nearest-neighbor 
$U_{02}(\omega)$), as shown by the dashed lines. 
 One can imagine that the variations in the effective Coulomb interaction at low frequencies will average out, so that the static limit $U(\omega=0)$ may still be appropriate for model Hamiltonian studies.


\section{\label{sec:conclusion}Conclusions}

In this work, we have employed the random phase approximation within the FLAPW method 
to study the on-site and $r$-dependent screening of the Coulomb interactions $U(r)$ 
(partially screened) and $W(r)$ (fully screened) in 2D TMDs MX$_2$ (M=Mo, W, Nb, 
Ta; X = S, Se, Te). Our results show that the $r$-dependent screening in semiconducting 
compounds like MoS$_2$ deviates substantially from the conventional behavior, 
i.e., $U(r)$ and $W(r)$ cannot be expressed by a simple static dielectric constant 
$\epsilon$. We found that the short-range interactions are strongly screened, 
while the long-range interactions are weakly screened, i.e., they decay much 
slower than the bare $1/r$ interaction. This nonconventional screening 
of the Coulomb interaction in 2D TMDs can be attributed to the reduced dimensionality 
of these materials. This $r$-dependent nonconventional screening explains the large 
exciton binding energies and deviations from the usual hydrogenic Rydberg series 
of energy levels of the excitonic states in semiconducting monolayer TMDs.

Our results also show that metallic TMDs like NbS$_2$ in the $1H$ structure 
exhibit a correlation strength $U/W_b \sim 2$, which is significantly larger 
than the corresponding values in elementary TMs. This is due to the strong 
$r$-dependent screening, which leads to a larger effective Coulomb interaction. 
The large $U/W_b$ ratio suggests that these materials are prone to Mott insulating 
behavior, which has been experimentally observed in the reconstructed $1T$ 
lattice. Using the calculated $U$ parameters for undistorted $1T$ 
structure, we extract the on-site effective $U^{\mathrm{eff}}_{00}$ and 
nearest neighbor $U^{\mathrm{eff}}_{01}$ Coulomb interaction parameters for 
reconstructed CCDW NbX$_2$ and TaX$_2$ compounds. Strictly speaking, for 
the reconstructed star of David $1T$-MX$_2$ (M=Nb, Ta) structure, the 
large $U^{\mathrm{eff}}_{00} \sim 0.8 $ eV estimated for $4d/5d$ electrons compared with 
the relatively small bandwidth $W_{b} \sim $ 0.02 eV satisfies the condition 
of $U^{\mathrm{eff}}_{00}/W_b \gg 1$. Furthermore, we found that the long-range 
Coulomb interactions remain sizable in metallic TMDs, despite the metallic nature of these 
materials. This is due to the reduced screening at long distances. 
The long-range Coulomb interactions can lead to the existence of intriguing 
plasmons in the monolayer of TMDs and the appearance of CDW order.

This study presents a comprehensive understanding of the Coulomb interactions in 
both semiconducting and metallic 2D TMDs. Furthermore, the derived Coulomb interaction 
parameters can be effectively employed in model Hamiltonians and DFT+$U$ (DFT+$U$+$V$) methods, 
consequently boosting the predictive power of these techniques.


\acknowledgements
E.\c{S} and I.M. acknowledge support from \textit{Sonderforschungsbereich} TRR 227 of 
Deutsche Forschungsgemeinschaft (DFG) and funding provided by the European Union
(EFRE), Grant No: ZS/2016/06/79307.

\bibliography{my-bib}
\end{document}